\documentclass[12pt]{iopart}

\expandafter\let\csname equation*\endcsname\relax
\expandafter\let\csname endequation*\endcsname\relax

\usepackage[utf8]{inputenc}
\usepackage{iopams}
\usepackage{amsmath}
\usepackage{amsthm}
\usepackage{amssymb}
\usepackage{breqn}
\usepackage{caption}
\usepackage[british]{babel}
\usepackage{graphicx} 
\usepackage{color}
\usepackage[colorlinks=true,linkcolor=blue,citecolor=blue]{hyperref}
\usepackage{mathtools}
\usepackage[T1]{fontenc}
\usepackage{cite}
\usepackage{mathrsfs}
\usepackage{subfigure}
\usepackage{float}

\begin{document}

\title{Exactly solvable diffusions from space-time transformations}

\author{Costantino Di Bello$^\dagger$, \'Edgar Rold\'an$^{\S}$, and Ralf
Metzler$^{\dagger\sharp}$}
\address{$\dagger$ University of Potsdam, Institute of Physics \& Astronomy,
14476 Potsdam, Germany\\
$\S$ ICTP---The Abdus Salam International Centre for Theoretical Physics,
34151 Trieste, Italy.\\
$\sharp$ Asia Pacific Centre for Theoretical Physics, Pohang 37673,
Republic of Korea}
\ead{costantinodibello@hotmail.it; edgar@ictp.it;
rmetzler@uni-potsdam.de (Corresponding author: Ralf Metzler)}

\begin{abstract}
We consider a general one-dimensional overdamped diffusion model described by
the It\^{o} stochastic differential equation (SDE) ${dX_t=\mu(X_t,t)dt+\sigma
(X_t,t)dW_t}$, where $W_t$ is the standard Wiener process. We obtain a specific
condition that $\mu$ and $\sigma$ must fulfil in order to be able to solve the
SDE via mapping the generic process, using a suitable space-time transformation,
onto the simpler Wiener process. By taking advantage of this transformation, we
obtain the propagator in the case of open, reflecting, and absorbing \emph{
time-dependent\/} boundary conditions for a large class of diffusion processes.
In particular, this allows us to derive the first-passage time statistics of
such a large class of models, some of which were so far unknown. While our
results are valid for a wide range of non-autonomous, non-linear and
non-homogeneous processes, we illustrate applications in stochastic
thermodynamics by focusing on the propagator and first-passage-time statistics
of isoentropic processes that were previously realized in the laboratory  with Brownian particles trapped with optical tweezers.
\end{abstract}

\section{Introduction}
\label{sec:intro}

The effort to understand the fundamental nature of reality is a timeless
pursuit that has fascinated humanity for centuries. One of the first proofs in
support of the atomistic theory of matter is the famous paper by Einstein on
Brownian motion \cite{Einstein}. Contemporarily, Sutherland \cite{sutherland},
Smoluchowski \cite{smoluchowski}, and Langevin \cite{langevin} all contributed
towards the physical theory of Brownian motion. The successive formalisation
by Wiener \cite{wiener} established the foundations of non-equilibrium
statistical mechanics. The partial differential equations describing
stochastic motion were developed independently by Fokker and Planck
\cite{fokker,planck} and by Kolmogorov \cite{kolmogorov1938} (see also
\cite{levy,coffey}). Indeed, in modern literature they either go under the name
of \emph{forward and backward Fokker-Planck\/} equations, or \emph{first
and second Kolmogorov\/} equations. An alternative mathematical formulation
of diffusion processes is via stochastic differential equations, firstly
developed by It\^{o}\cite{Ito1950}. Nevertheless, stochastic processes
have been applied not only to the Brownian motion but have found numerous
applications in various fields. For instance, to study the motion of passive
molecules \cite{Hofling_2013,arbel-goren2023,lene,garini} and actively
transported particles \cite{jaenatc,christine} in biological cells,
lipids in membranes \cite{jaememb,mattijcp}, animal
motion \cite{animal_motion}, active matter \cite{RevModPhys_active}, geophysics
\cite{BERKOWITZ2002861}, condensed matter \cite{Scher_1975}, financial markets
\cite{Bouchaud_Potters_2003}, or disease spreading \cite{brockmann_network}.
One of the most significant random variables for stochastic models is the
first-passage time (FPT), a positive random variable defined as the first time
at which a stochastic walker reaches a threshold or exits from a certain region
of space \cite{Redner_2001,ralf_book_FPT}. To calculate the full distribution
of FPTs even in simple geometries is a formidable task, that can be solved by,
e.g., methods such as Newton series, spectral methods or self-consistent
approaches \cite{aljaz,aljaz1,denis}. One of the most prominent examples
is the Kramers problem \cite{kramers,hanggi_review}, widely used to model
the activation rate in chemical reactions. Applications of the FPT are
extensive and a complete literature report would be a formidable task. As a
few examples, we mention applications to animal foraging \cite{Viswanathan},
to model the disease spreading of infections \cite{hufnagel_pnas,Gross_2020},
or in finance \cite{valenti_pre,valenti_ijbc} where information regarding
the FPT is fundamental to determine actions such as buying or selling. The
FPT also provides valuable information regarding the extreme values of
random processes \cite{Hartich_jpa} and on observables in non-equilibrium
statistical physics \cite{Bray_majumdar}. A recent survey on applications
is available here\cite{target_search_problems}.

In this paper, we address the problem of solving stochastic differential
equations (SDEs) (or Fokker-Planck equations) and FPT problems
via a space-time coordinate transformation. Further details on the
physical interpretation of this transformation are presented in section
\ref{Sec:interpretation}. Nevertheless, as a similar idea is roughly
outlined in some old works, we first provide a literature overview of
these references. Probably the first author to have the idea of space-time
transformation was Kolmogorov \cite{kolmogorov1938}. In section 17 of his
seminal work, Kolmogorov discussed a couple of examples of Fokker-Planck
equations (FPEs) that can be solved by a change of variables, i.e.,
transforming $x\to x'$ and $t\to t'$. Some 25 years later, Cherkasov
\cite{cherkasov1957}, another Russian mathematician, following the same idea
of variable change, discovered a sufficient and necessary condition to map
a generic FPE onto the FPE of the Wiener process. Nevertheless, Cherkasov's
work showed an error in the proof, which was later corrected by Shirokov
\cite{cherkasov_correction}.  Interestingly, the paper \cite{cherkasov1957}
was nearly forgotten. As Ricciardi, an Italian mathematician, wrote in
his paper \cite{ricciardi1976} "{\em apparently Cherkasov's work did not
receive much attention in the Western world}". In \cite{ricciardi1976} the
author basically reproduces the results of \cite{cherkasov1957}, yet with a
clearer notation.  Later on, Ricciardi and collaborators \cite{ricciardi1984}
mentioned, without proof, that a similar mapping technique could be adapted to
solve FPT problems. A similar idea can be found in \cite{wang}. We credit the
authors of \cite{ricciardi1984} for their intuition, although, to the authors'
knowledge, this idea has never been used by any author to solve FPT problems.

The aim of our paper is to both present some of the results of the
aforementioned papers in a simpler way and both to further develop some of
the ideas and to obtain the first-passage time density (FPTD) analytically for
a large class of diffusion models. Therefore, throughout the text, we will
expose a mixture of known and novel results. We hope that this paper will serve
as a key reference to the physics community for these valuable yet arguably
forgotten transformation techniques. As a particular example of application,
we discuss (i) the isoentropic protocol, an important
diffusion process
for stochastic thermodynamics and related experiments~\cite{Schmiedl, martinez}; and
(ii) the stochastic Gompertz model, used in population dynamics (see~\cite{ricciardi1999} for a review). The paper
is organised as follows: in section \ref{sec:main_results} we outline, in
four different
subsections, the main results of the paper. In sections \ref{Sec:interpretation}
and \ref{sec:FPT} we provide an intuitive explanation of our technique,
respectively for the propagator and the FPT. In section \ref{sec:examples}
we focus on specific applications of our theory, and section \ref{exist}
details some connections with existing results. Our Conclusions are drawn
in section \ref{concl}.

\section{Summary of the main results}
\label{sec:main_results}

\subsection{Solution of the SDE and propagator}
\label{sec:main_results1}

We start by defining the problem and the most important quantities needed
for presenting the results. We consider generic diffusion processes described
by the one-dimensional It\^{o} stochastic differential equation (SDE)
\cite{oksendal}
\begin{equation}
\label{eqn:Ito_SDE}
dX_t=\mu(X_t,t)dt+\sigma(X_t,t)dW_t,
\end{equation}
with initial condition $X_{t_0}=x_0$, where $W_t$ is the standard Wiener
process. 

Let us introduce the following function, which will be crucial later on,
\begin{equation}
\label{eqn:cherkasov}
\mathcal{C}(x,t)\equiv\dfrac{1}{\sigma(x,t)}\dfrac{\partial\sigma(x,t)}{\partial
t}+\sigma(x,t)\dfrac{\partial}{\partial x}\left(\dfrac{1}{2}\dfrac{\partial
\sigma(x,t)}{\partial x}-\dfrac{\mu(x,t)}{\sigma(x,t)}\right).
\end{equation}
Here and below, we call $\mathcal{C}(x,t)$ the {\em Cherkasov function}. Notice
that $\mathcal{C}(t)$ has physical dimensions of inverse time. Although the
function $\mathcal{C}$ never appears in the works of Cherkasov, we decided to
use the notion "Cherkasov function" in Cherkasov's honour. As we show later
in~\ref{appendix1}, if $\mathcal{C}(x,t)\equiv\mathcal{C}(t)$ is
solely a function of time, i.e.,
\begin{equation}
\label{eqn:condition}
\dfrac{\partial\mathcal{C}(x,t)}{\partial x}=0,
\end{equation}
then it is possible to obtain an exact analytical solution of the SDE
\eqref{eqn:Ito_SDE} and an exact expression for the propagator of the
process. More precisely, when equation \eqref{eqn:condition} holds, then it is
possible to find two real {\em deterministic\/} functions $\psi(x,t)$
[equation \eqref{eqn:psi}] and $\tau(t)$ [equation \eqref{eqn:tau}], both
invertible, that enable one to solve explicitly the SDE \eqref{eqn:Ito_SDE} as
\begin{equation}
\label{eqn:SDE_solution}
X_t=\psi^{-1}(W_{\tau(t)},t);
\end{equation}
in other words, $W_{\tau(t)}=\psi(X_t,t)$, i.e., the  time  reparametrisation
of the Wiener process equals the function $\psi$ evaluated along the process
$X_t$. Note that the time change $t\to\tau$ is not a random-time transformation
but a deterministic one-to-one mapping, yet $W_{\tau(t)}$ is a martingale
\cite{martingale_review}.
Equivalently to \eqref{eqn:Ito_SDE}, we can say that
the propagator of the process, $P(x,t|x_0,t_0)dx\equiv\mathrm{Prob}\{x<X_t=x+
dx|X_{t_0}=x_0\}$ fulfils the Fokker-Planck equation
\begin{equation}
\label{eqn:FPE}
\dfrac{\partial P(x,t|x_0,t_0)}{\partial t}=\dfrac{1}{2}\dfrac{\partial^2}{
\partial x^2}\left(\sigma^2(x,t)P(x,t|x_0,t_0)\right)-\dfrac{\partial}{\partial
x}\left(\mu(x,t)P(x,t|x_0,t_0)\right),
\end{equation}
with delta initial condition $\lim_{t\to t_0}P(x,t|x_0,t_0)=\delta(x-x_0)$ and
with some appropriate boundary conditions (BC). We remind the reader that the
probability flux associated with the FPE \eqref{eqn:FPE} obeying $\partial_t
P(x,t|x_0,t_0)=-\partial_x j(x,t|x_0,0)$ is defined as 
\begin{equation}
j(x,t|x_0,t_0)\equiv\mu(x,t)P(x,t|x_0,t_0)-\dfrac{1}{2} \dfrac{\partial}{
\partial x}\left(\sigma^2(x,t) P(x,t|x_0,t_0)\right).
\end{equation}
 
Property \eqref{eqn:SDE_solution} implies that the propagator of the process
for natural
(open) boundary conditions ({${\lim_{|x|\to\infty}P(x,t|x_0,t_0)=0}$})
reads
\begin{equation}
\label{eqn:propagator}
P(x,t|x_0,t_0)=\dfrac{\partial\psi(x,t|x_0,t_0)}{\partial x}\dfrac{1}{\sqrt{2
\pi\tau(t|t_0)}}\exp\left(-\dfrac{\psi^2(x,t|x_0,t_0)}{2\tau(t|t_0)}\right).
\end{equation}
The functions $\psi(x,t|x_0,t_0)$ and $\tau(t|t_0)$, which have dimensions of
square root of time and time, respectively, are given by
\begin{eqnarray}
\nonumber
\psi(x,t|x_0,t_0)&\equiv&\exp\left(\int_{t_0}^t\mathcal{C}(s)ds\right)\int_{
x_0}^x\dfrac{1}{\sigma(x',t)}dx'\\
&&+\int_{t_0}^tds\left(\dfrac{1}{2}\dfrac{\partial\sigma}{\partial x} \bigg|_{
(x_0,s)}-\dfrac{\mu(x_0,s)}{\sigma(x_0,s)}\right)\exp\left(\int_{t_0}^s\mathcal{
C}(s')ds'\right)
\label{eqn:psi}
\end{eqnarray}
and
\begin{equation}
\label{eqn:tau}
\tau(t|t_0)\equiv\int_{t_0}^t\exp\left(2\int_{t_0}^s\mathcal{C}(s')ds'\right)ds.
\end{equation}
Both functions $\psi$ and $\tau$ depend parametrically on  $x$ and $t$ and their initial conditions
$x_0$ and $t_0$ (through the functions $\mu$, $\sigma$ and $\mathcal{C}$ which is given by Eq.~\eqref{eqn:cherkasov}), however, we will omit this dependence in the notation throughout
the manuscript whenever there is no risk of confusion. The proof of formulae
\eqref{eqn:SDE_solution}, \eqref{eqn:propagator}, \eqref{eqn:psi}, and
\eqref{eqn:tau} is shown in \ref{appendix1}; they are the first main result of
the paper. We obtained formulae \eqref{eqn:propagator}, \eqref{eqn:psi}, and
\eqref{eqn:tau} with a significantly simpler proof as compared to the original
work \cite{cherkasov1957}, while \eqref{eqn:SDE_solution} is a novel result of
this paper. Condition \eqref{eqn:condition} was also found in \cite{wang}.
It remains unclear, to the authors' knowledge (and an
interesting avenue for future research), whether condition
\eqref{eqn:condition} can be understood from intuitive physical arguments.

The class of Wiener transformable SDEs and their propagators is constrained
by fulfilling equation \eqref{eqn:condition}. Nonetheless, there are examples of SDEs that, despite not fulfilling \eqref{eqn:condition} are solvable analytically~\cite{bray_log, giampaoli_anharmonic, ryabov_log-harmonic, skiadas}.
We noticed that the cases discussed in the aforementioned papers are equivalent to the so-called inhomogeneous geometric Brownian motion (IGBM) (see also~\cite{tubikanec}), which is discussed in 
subsection~\ref{sec:second_cherkasov}. 
Moreover, within the class defined by~\eqref{eqn:condition}, not all the
SDEs are amenable to analytical solution for their FPTD. In the next section
we outline for which problems we can obtain exact closed forms for the FPTD.

\subsection{First-passage time density}
\label{sec:main_results2}

Following the discussion of the previous subsection, consider a generic
stochastic process described by equation \eqref{eqn:Ito_SDE} starting at $X_{t_0}
=x_0$ and with the boundary $a(t)$ with $a(t_0)>x_0$, we define the random
variable FPT $\mathcal{T}_X$ as
\begin{equation}
\label{eqn:FPT_def}
\mathcal{T}_X(a(t)|x_0,t_0)\equiv\inf\left\{ t>0:X_t>a(t)|X_{t_0}=x_0\right\}.
\end{equation}
In simple words this represents the first time that the stochastic process $X_t$, which started below $a(t)$, crosses the boundary.
We specify that the boundary $a(t)$ may be any time-dependent yet deterministic
function of time $t$. To avoid problems in this definition \eqref{eqn:FPT_def}
we further assume that $a(t)$ is a continuous function. The definition in the
case $a(t_0)<x_0$ is analogous. The analytical result we obtained on the FPTD,
besides expression \eqref{eqn:condition}, further requires another condition,
namely, that the FPT of the transformed process $W_\tau$ to reach the transformed
threshold $\psi(a(t),t)$ is analytically solvable. An example class for which
this is possible is when
\begin{equation}
\label{eqn:condition2}
\psi(a(t),t)=v\tau(t)+a_0,
\end{equation}
with $v,a_0\in\mathbb{R}$ being constants, that is, $\psi$ evaluated at the
boundary depends linearly on $\tau$. We denote by $\wp(a(t),t|x_0,t_0)dt\equiv
{\rm Prob}\left\{t<\mathcal{T}_X(a(t)|x_0,t_0)<t+dt\right\}$ 
the FPTD. Assuming that \eqref{eqn:condition} and
\eqref{eqn:condition2} hold, it reads
\begin{equation}
\label{eqn:FPT_distr}
\wp(a(t),t|x_0,t_0)=\dfrac{|a_0|}{\sqrt{2\pi\tau(t)^3}}\exp\left(-\dfrac{
(a_0+v\tau(t))^2}{2\tau(t)}\right)\left(\dfrac{d\tau(t)}{dt}\right). 
\end{equation}
Formulae \eqref{eqn:condition2} and \eqref{eqn:FPT_distr} are the second main
result of the paper. We stress the fact that formula \eqref{eqn:FPT_distr} is
a generalisation of several results already known in the literature, such as
those reported in \cite{molini,grebenkov}. Further results for absorbing and
reflecting boundaries are given in the next subsection.

\begin{figure}
\centering
\includegraphics[width=0.4\linewidth]{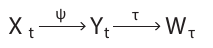}\\[0.6cm]
\includegraphics[width=1.0\linewidth]{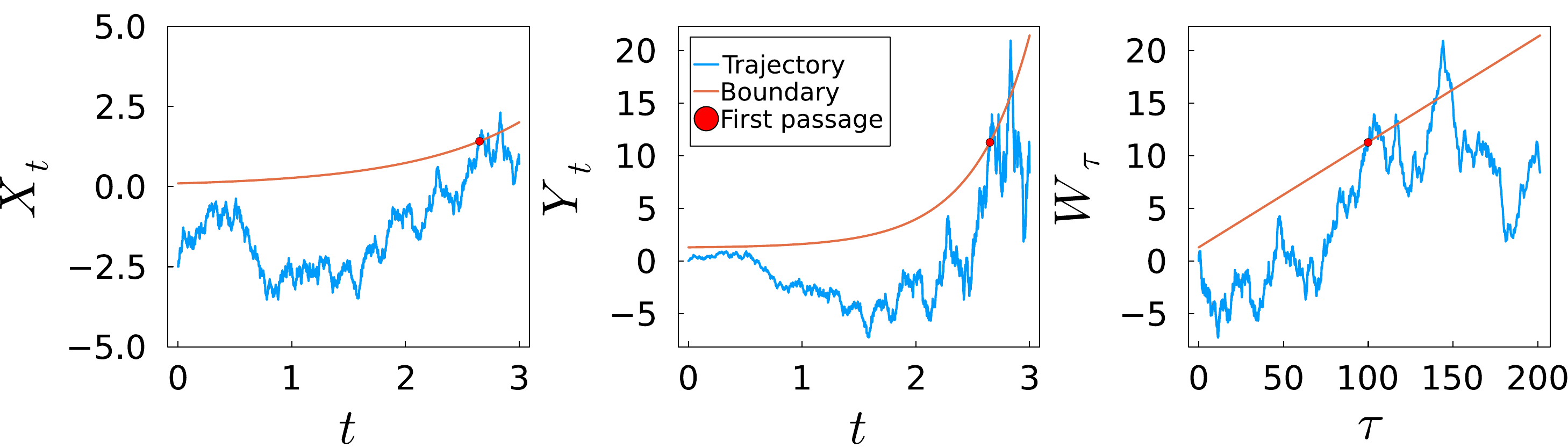}
(a) Ornstein-Uhlenbeck (OU) process\\[0.8cm]
\includegraphics[width=1.0\linewidth]{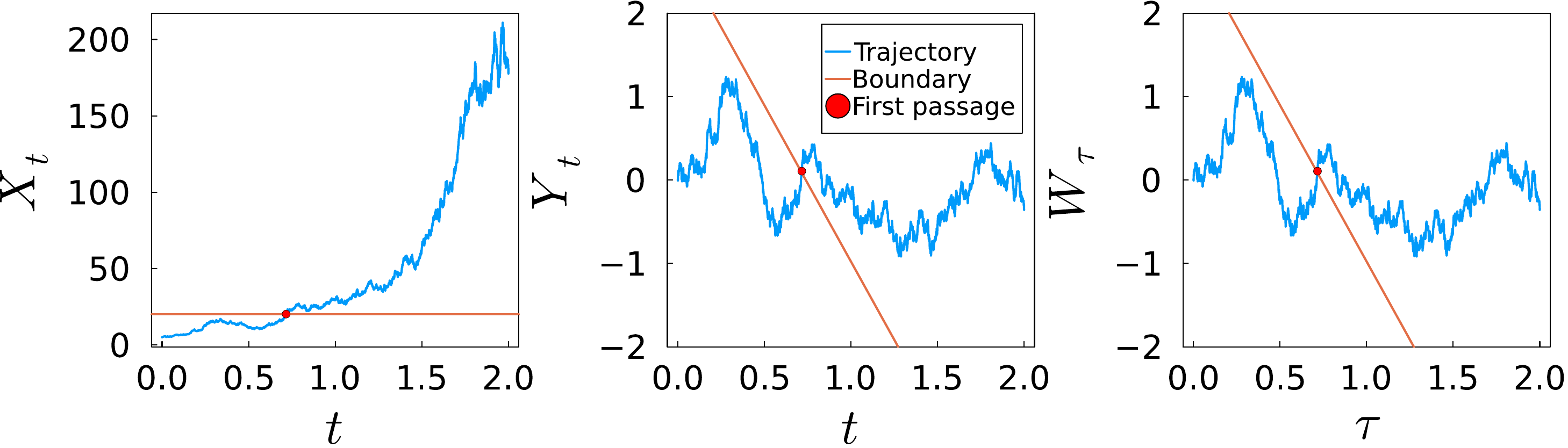}
(b) Geometric Brownian motion (GBM)
\caption{Examples of spatio-temporal transformations of stochastic processes
(top) that enable an analytical treatment of their propagator and FPT
statisticsusing the method of the current pape, along with illustrations for two examples: (a) Ornstein-Uhlenbeck
(OU) process ($dX_t=-\mu X_tdt+\sigma dW_t$, blue line) and FPT to reach a
time-dependent boundary (red line). (b) Geometric Brownian motion (GBM) ($dX_t
=\mu X_tdt+\sigma X_tdW_t$, blue line) and FPT to reach a constant boundary. In
the left panels we show the original processes $X_t$, in the central panels the
space-transformed processes $Y_t=\psi(X_t,t)$, while in the right panels the
space-time-transformed processes $W_\tau$ (Wiener) are shown. In (a) we
considered a time-dependent boundary that grows exponentially with time $a(t)
=a(0)e^{\mu t}$, while in (b) we have a constant boundary. In both cases, after
the transformation, the boundary becomes linear in time. Note that for GBM, the
time transformation is trivially $\tau=t$ (cf.~\eqref{eqn:tau}). The moment of
the first passage is highlighted by the red circle in all panels.
\label{fig:transformations}}
\end{figure}

\subsection{Propagators for absorbing and reflecting boundary conditions}
\label{sec:main_results3}

If the two conditions \eqref{eqn:condition} and \eqref{eqn:condition2} are met,
further results are available for the two cases with (i) {\em absorbing boundary
condition\/} $P_a(a(t),t|x_0,t_0)=0$; and (ii) {\em reflecting boundary
condition\/} (see \ref{appendix2} for a proof)
\begin{equation}
\label{eq:RBC}
j_r(a(t),t|x_0,t_0)=a'(t)P_r(a(t),t|x_0,t_0),
\end{equation}
where $j_r$ represents the probability flux at the time-dependent position $a(t)$ of the boundary.
If conditions \eqref{eqn:condition} and
\eqref{eqn:condition2} are valid, the two propagators for absorbing and
reflecting boundary, respectively $P_a$ and $P_r$, are known analytically
and read
\begin{equation}
\label{eqn:propagator_abs}
P_a(x,t|x_0,t_0)=\dfrac{\partial\psi}{\partial x}\frac{1}{\sqrt{2\pi\tau}}
\left[\exp\left(-\frac{\psi^2}{2\tau}\right)-\exp\left(-2a_0v-\frac{(\psi
-2a_0)^2}{2\tau}\right)\right]
\end{equation}
and
\begin{eqnarray}
\nonumber
P_r(x,t|x_0,t_0)&=&\dfrac{\partial\psi}{\partial x}\left\{\frac{1}{\sqrt{2\pi
\tau}}\left[\exp\left(-\frac{\psi^2}{2\tau}\right)+\exp\left(-2a_0v-\frac{
\left(\psi-2a_0\right)^2}{2\tau}\right)\right]\right.\\
&&\left.-v\exp\left(-2v(\psi-a_0-v\tau)\right){\rm erfc}\left(\frac{\psi-2a_0
-2v\tau}{\sqrt{2\tau}}\right)\right\},
\label{eqn:propagator_ref}
\end{eqnarray}
where $\mathrm{erfc}$ is the complementary error function. To make the
equations shorter, we omitted the explicit dependencies of the functions
$\psi$ and $\tau$. Equations \eqref{eqn:propagator_abs} and
\eqref{eqn:propagator_ref} are the third main result of the paper.
The proof of the results for these two cases is found in \ref{appendix2}.
We verified that these formulae solve the FPE with the appropriate boundary
condition with Mathematica.

\subsection{Solvable yet non-transformable SDEs and second Cherkasov condition}
\label{sec:second_cherkasov}
As stated at the end of subsection \ref{sec:main_results1},
there are some SDEs that are solvable even though they do not
fulfil condition~\eqref{eqn:condition}. These cases are discussed in
Refs.~\cite{bray_log, giampaoli_anharmonic, ryabov_log-harmonic, skiadas,
tubikanec} and interestingly they are all equivalent to the
so-called {\em inhomogeneous geometric Brownian motion} (IGBM)~\cite{skiadas}. The SDE for IGBM reads
\begin{equation}
\label{eqn:IGBM}
dX_t=\left(\alpha(t) X_t+\beta(t)\right)dt+\bar{\sigma}(t)X_tdW_t,
\end{equation}
where $\alpha$, $\beta$, and $\bar{\sigma}$ may be any functions of time. Thus,
the only difference with respect to GBM is the presence of $\beta(t)dt$ in the
SDE~\eqref{eqn:IGBM}.
Therefore, we may argue that there
is a second class of SDEs that, even though they cannot be transformed into the
Wiener process, are instead mappable onto IGBM The coefficients of such SDEs,
while not fulfilling \eqref{eqn:condition}, satisfy the condition
\begin{equation}
\label{eqn:condition3}
\dfrac{\partial}{\partial x}\left[\left(\dfrac{\partial\mathcal{C}(x,t)}{
\partial x}\right)^{-1}\dfrac{\partial}{\partial x}\left(\sigma(x,t)\dfrac{
\partial\mathcal{C}(x,t)}{\partial x}\right)\right]=0,
\end{equation}
where $\mathcal{C}$ was defined in~\eqref{eqn:cherkasov}. Moreover, the
transformation $\psi$ that maps the original process onto IGBM reads
\begin{equation}
\psi(x,t) = \left( \sigma(x,t) \dfrac{\partial
\mathcal{C}(x,t)}{\partial x} \right)^{-1}
\end{equation}
We here do not report the solution of equation \eqref{eqn:IGBM} as it is available
in standard textbooks \cite{Mao_book}. The proof of these results is provided
in \ref{app:second_cherkasov}.

\section{Interpretation of the results: mapping a stochastic process onto
the Wiener process}
\label{Sec:interpretation}

In this section we provide a more physical intuition on the meaning of the
main formulae \eqref{eqn:SDE_solution}, \eqref{eqn:propagator}, \eqref{eqn:psi},
and \eqref{eqn:tau}. A graphical representation of our discussion is shown in
Fig.~\ref{fig:transformations}. We will not expose the proof here, this is
available in \ref{appendix1}. Let us start again from the SDE (\ref{eqn:Ito_SDE}),
for which the coefficients $\mu(x,t)$ and $\sigma(x,t)$ go under several names:
in physics $\mu(x,t)$ represents a deterministic force acting on the particle,
while $\sigma(x,t)dW_t$ is the noise term due to thermal fluctuations. It is
related to the diffusivity of the process via $D(x,t)=\sigma^2(x,t)/2$. Other
popular names, especially in financial literature, for $\mu$ and $\sigma$ are
respectively the {\em drift\/} and the {\em volatility\/} (e.g., see
\cite{martingale_review}).

As we show in \ref{appendix1}, if and only if condition \eqref{eqn:condition}
holds, i.e., if and only if $\mathcal{C}(t)$ is solely a function of time, then
it is possible to define a new process $Y_t=\psi(X_t,t)$ (for simplicity we do
not state the dependencies on $x_0$ and $t_0$, as it is obvious) that the
simpler SDE describes
\begin{equation}
dY_t=\sigma_Y(t)dW_t,
\end{equation}
where the volatility $\sigma_Y(t)$ of the new process is a function of time,
only, and with $\psi(x,t)$ being strictly increasing, thus invertible, with
respect to the first variable. For simplicity, without loss of generality,
we assume that the initial condition of the new process is zero, $Y_{t_0}=
\psi(x_0,t_0)\equiv0$. The explicit form of $\psi(x,t)$ based on these
features is given in \eqref{eqn:psi} and derived in \ref{appendix1}. This
means that $Y_t$ is a time-transformed version of a Wiener process. Therefore,
we can define the new time variable 
\begin{equation}
\tau\equiv\int_{t_0}^t\sigma_Y^2(s)ds,
\end{equation}
such that the reindexed process $Y_{\tau}$ is just a Wiener process,
\begin{equation}
Y_{\tau}=W_{\tau},
\end{equation}
with initial condition $Y_{\tau=0}=0$. In other words, if and only if
condition \eqref{eqn:condition} holds, there exists a special framework of
coordinates, in which the stochastic motion is perceived as a simple Brownian
motion. Thus, the connection between $X_t$ and the Brownian motion $W_t$ is
\begin{equation}
W_{\tau(t)}=\psi(X_t, t),
\end{equation}
as stated in equation \eqref{eqn:SDE_solution}. Moreover, since the function $\psi
(x,t)$ is invertible (see \ref{appendix1}) it is possible to write
\begin{equation}
\label{eqn:psi-1}
X_t=\psi^{-1}(W_{\tau(t)},t),
\end{equation}
from which it is possible to obtain the correlations of the process as shown
for the specific example discussed in section \ref{sec:OU}.

We now consider the propagator $P_W(y,\tau|0,0)\equiv P_Y(y,\tau)$ of the process
$W_\tau$. It is connected to the sought-after propagator $P(x,t|x_0,t_0)$ via
the following change of measure (for consistency between the left and right
hand sides we explicitly include the dependencies on $x_0,t_0$)
\begin{equation}
\label{eqn:P_Pprime}
P(x,t|x_0,t_0)=\dfrac{\partial\psi(x,t|x_0,t_0)}{\partial x} P_W(y,\tau),
\end{equation}
where we note that this formula relies on the fact that $\partial\psi/\partial
x>0$. Using the fact that $P_W(y,\tau)$ is the propagator of a Brownian motion
and rewriting the right hand side of equation \eqref{eqn:P_Pprime} in terms of the
original variables $(x,t)$, we find
\begin{eqnarray}
\nonumber
P(x,t|x_0,t_0)&=&\dfrac{\partial\psi(x,t|x_0,t_0)}{\partial x}\dfrac{1}{\sqrt{2
\pi\tau}}\exp\left(-\dfrac{y^2}{2\tau}\right)\\
&=&\dfrac{\partial\psi(x,t|x_0,t_0)}{\partial x}\dfrac{1}{\sqrt{2\pi\tau(t|t_0
)}}\exp\left(-\dfrac{\psi^2(x,t|x_0,t_0)}{2\tau(t|t_0)}\right).
\end{eqnarray}
The explicit forms of $\psi(x,t|x_0,t_0)$ and $\tau(t|t_0)$ are shown in
equations \eqref{eqn:psi} and \eqref{eqn:tau}.

A graphical representation of our discussion is provided in figure
\ref{fig:transformations} where we also highlight how the FPT for a
specific boundary varies after the space-time transformations. We conclude
this section with the following observation: the idea of using space-time
transformations is in fact reminiscent of the Cameron-Martin-Girsanov
theorem \cite{liptser2013statistics}. This theorem describes how an SDE
can be simplified via a change of measure, after which the transformed
process becomes a martingale. While the Cameron-Martin-Girsanov theorem has
found prominent applications in diverse fields such as finance (e.g., the
celebrated Black-Scholes formalism \cite{merton1971theory}), it does not
provide a clear-cut criterion on which an SDE is amenable to exact
analytical expressions for their propagators and FPTDs.

\section{Which FPT problems are solvable?}
\label{sec:FPT}

We now turn our attention to the FPT problem. The question we are trying to
address is: supposing that condition \eqref{eqn:condition} is fulfilled,
is it possible to obtain the FPTD? Why do we also require condition
\eqref{eqn:condition2}? First of all, supposing that \eqref{eqn:condition}
is fulfilled, we could transform the stochastic process $X_t$ into the Wiener
process, and then consider the FPT problem for a Wiener process. In
mathematical terms the idea is the following,
\begin{eqnarray}
\nonumber
\inf\left\{ t>0:X_t>a|X_{t_0}=x_0\right\}&=&\inf\left\{ t>0:Y_t>\psi(a,t)|Y_{
t_0}=0\right\}\\
\nonumber
&=&\inf\left\{ t(\tau)>0:Y_{t(\tau)}>\psi(a,t(\tau))|Y_{t(0)}=0\right\}\\
&=&\tau^{-1}\left(\inf\left\{\tau>0:W_\tau>\psi(a,\tau)|W_0=0\right\}\right),
\end{eqnarray}
where in the first equality we used that the space-transformation $\psi$ is
monotonically increasing, in the second we expressed everything in terms of
$t\equiv\tau^{-1}$ since the function $\tau$ is invertible (see
\ref{appendix1}), and in the third one we expressed everything in terms of
$\tau$. We also used the fact that $Y_\tau=W_\tau$. By denoting the FPT of
a Wiener process with $\mathcal{T}_W$, we thus proved that
\begin{equation}
\label{eqn:relation_FPTs}
\mathcal{T}_W(\psi(a,t))=\tau\left(\mathcal{T}_X(a|x_0,t_0)\right),
\end{equation}
where on the right hand side we dropped the dependence on initial position and
time as both are $0$.

Therefore the FPT problem of the original process $X_t$ for the barrier $a(t)$
is mapped onto the FPT of the Wiener process for the barrier $\psi(a(t),t)$
(refer to Fig~\ref{fig:transformations} for a geometric intuition). An explicit
pedagogical explanation of this transformation is available in
\ref{appendix_didactic} for the case of the Ornstein-Uhlenbeck process. We
remind the reader that the function $\tau(t|t_0)$ defined in \eqref{eqn:tau}
is strictly increasing. In terms of probability distributions,
equation \eqref{eqn:relation_FPTs} reads
\begin{equation}
\label{eqn:FPTD_rel}
\wp_X(a(t),t|x_0,t_0)=\wp_W(\psi(a(t),t),\tau(t))\dfrac{d\tau(t|t_0)}{dt},
\end{equation}
where $\wp_W$ is the FPTD of the Wiener process. Analytical results for $\wp$
are available only for a few limited cases.

The first-passage or first-crossing problem for the Wiener process to a
time-dependent yet deterministic boundary is a long-standing open question
in mathematics, several authors tried approached this problem with different
analytical and computational methods \cite{robbins_siegmund1970,siegmund1986,
ricciardi1984,wang,durbin,alili2010}. Nevertheless, complete analytical
results are available only for certain types of boundaries, such as those with
linear dependencies \cite{lerche}, as stated in equation \eqref{eqn:condition2}.
Either way, even though the transformed boundary is
non-linear, transforming the original stochastic process into the Wiener process
is advantageous to find the FPTD, since in this case approximation schemes
are available~\cite{wang, wang97}. Due to technical difficulties, we here do
not report the case of boundaries with a nonlinear relation between $\psi$
and $\tau$, which we reserve for a future publication. Substituting into
equation \eqref{eqn:FPTD_rel} the explicit form of the FPTD of the Wiener process
(Wald or L\'evy-Smirnov distribution) for this transformed linear boundary
$\psi(a(t),t)$, we finally complete the proof of equation \eqref{eqn:FPT_distr} (see
\cite{coxbook,goelbook} for the derivation of the FPTD of the Wiener process).

Analogously, the same argument applies to propagators with absorbing and
reflecting boundaries. As the FPTD, they are well known analytically only
for certain boundaries. Further details on the proof of formulae
\eqref{eqn:propagator_abs} and \eqref{eqn:propagator_ref} can be found in
\ref{appendix2}. In \ref{appendix_didactic} it is shown how the FPTD of an
Ornstein-Uhlenbeck process for a constant boundary can be mapped onto the
FPTD of the Wiener process for a square-root boundary. We refer to
\cite{breiman} for analytical results on square-root boundaries.

\section{A few examples of application}
\label{sec:examples}

The main results of this paper, equations \eqref{eqn:SDE_solution},
\eqref{eqn:propagator}-\eqref{eqn:tau}, \eqref{eqn:FPT_distr},
\eqref{eqn:propagator_abs}, and \eqref{eqn:propagator_ref} can provide
several results for many stochastic processes that so far were believed
to be impossible to tackle analytically. As specific simple examples, we
selected the (i) Ornstein-Uhlenbeck process with time-dependent stiffness
and temperature, and (ii) the isoentropic protocol. We proceed with a
discussion of these examples.

\subsection{Non-autonomous Ornstein-Uhlenbeck process}
\label{sec:OU}

Recent experiments explored heat engines, in which colloidal particles subject
to a time-dependent temperature are confined by optical tweezers with
time-dependent stiffness~\cite{ciliberto2017experiments,martinez2017colloidal}.
The overdamped Langevin equation describing the fluctuating motion of the
position of the  particle is
\begin{equation}
\label{eqn:langevin}
dX_t=-\dfrac{\kappa(t)}{\gamma}X_tdt+\sqrt{\dfrac{2k_BT(t)}{\gamma}}dW_t,
\end{equation}
where $\gamma$ is the friction coefficient, $\kappa(t)$ is the trap stiffness,
and $T(t)$ is the temperature. Both $\kappa(t)$ and $T(t)$ have a specific time
dependence as they change during the cycle. Despite the caveat issued in
\cite{frydel} for the explicit time dependence of these parameters, we stress
that the solution \eqref{eqn:propagator} is valid for {\em any\/} protocol
$\kappa(t)$, $T(t)$ driving equation \eqref{eqn:langevin}, as we discuss below.
In fact, the Cherkasov function \eqref{eqn:cherkasov} associated with this
non-autonomous process reads
\begin{equation}
\mathcal{C}(t)=\dfrac{1}{2}\dfrac{d\ln(T(t))}{dt}+\dfrac{\kappa(t)}{\gamma},
\end{equation}
which clearly fulfils condition \eqref{eqn:condition}. The integral of the
Cherkasov function reads
\begin{equation}
\int_0^t\mathcal{C}(s)ds=\ln\left(\sqrt{\dfrac{T(t)}{T(0)}}\right)+\Omega(t),
\end{equation}
where we introduced the time-integrated corner frequency as
\begin{equation}
\Omega(t)\equiv\int_0^t\frac{\kappa(s)}{\gamma}ds,
\end{equation}
which is a dimensionless quantity. We can compute the form of $\psi$ and
$\tau$ using equations \eqref{eqn:psi} and \eqref{eqn:tau}, yielding
\begin{equation}
\psi(x,t|x_0)=\sqrt{\dfrac{\gamma}{2k_BT(0)}}\left(xe^{\Omega(t)}-x_0\right)
\end{equation}
and
\begin{equation}
\tau(t)=\dfrac{1}{T(0)}\int_0^tT(s)e^{2\Omega(s)}ds.
\end{equation}
Thus, using expression \eqref{eqn:propagator}, the propagator becomes
\begin{eqnarray}
\nonumber
P(x,t|x_0,0)&=&\sqrt{\dfrac{\gamma}{2k_BT(0)}}\dfrac{1}{\sqrt{2\pi\int_0^t\frac{
T(s)}{T(0)}\exp\left(-2(\Omega(t)-\Omega(s))\right)ds}}\\
&&\times\exp\left(-\dfrac{\gamma}{2k_BT(0)}\dfrac{(x-x_0e^{-\Omega(t)})^2}{2
\int_0^t\frac{T(s)}{T(0)}e^{-2(\Omega(t)-\Omega(s))}ds}\right).
\label{eqn:propagator_OU}
\end{eqnarray}

Furthermore, as mentioned before, we can use formula \eqref{eqn:psi-1} to
compute the correlations of the process. For this aim we need the inverse
function of $\psi(x,t)$, which, in this case, reads
\begin{equation}
\psi^{-1}(x,t|x_0)=\left(\left[\sqrt{\dfrac{\gamma}{2k_BT(0)}}\right]^{-1}x+
x_0\right)e^{-\Omega(t)}.
\end{equation}
Therefore,
\begin{equation}
 X_t=\left(\left[\sqrt{\dfrac{\gamma}{2k_BT(0)}}\right]^{-1}W_{\tau(t)}+x_0
\right)e^{-\Omega(t)}
\end{equation}
and
\begin{equation}
\langle X_t X_{t'}\rangle=\left<\left(\sqrt{\dfrac{\gamma}{2k_BT(0)}}\right)
^{-2}W_{\tau(t)}W_{\tau(t')}+x_0^2\right> e^{-\Omega(t)-\Omega(t')},
\end{equation}
where we already removed all vanishing expectations. Keeping in mind that both
$\tau$ and $\Omega$ are strictly increasing functions, we get 
\begin{eqnarray}
\nonumber
\langle X_t X_{t'}\rangle&=&x_0^2e^{-\Omega(t)-\Omega(t')}+\left(\sqrt{\dfrac{
\gamma}{2k_BT(0)}}\right)^{-2}\tau\left(\min(t,t')\right)e^{-2\Omega\left(\min
(t,t')\right)}\\
&=&x_0^2e^{-\Omega(t)-\Omega(t')}+\left(\sqrt{\dfrac{\gamma}{2k_BT(0)}}\right)
^{-2}\int_0^{\min(t,t')}\dfrac{T(s)}{T(0)}e^{-2[\Omega\left(\min(t,t')\right)
-\Omega(s)]}ds.
\end{eqnarray}

Let us now turn to the FPTD. According to equation \eqref{eqn:condition2} it is
possible to compute it when the transformed boundary is linear in $\tau$.
Condition \eqref{eqn:condition2} can be reformulated in the equivalent form
\begin{equation}
\left.\dfrac{d\psi}{d\tau}\right|_{x=a(t)}=\mathrm{const}.
\end{equation}
In this case, $d\psi/d\tau $ reads
\begin{equation}
\label{eqn:dpsi_dtau_OU}
\left.\dfrac{d\psi}{d\tau}\right|_{x=a(t)}=\left.\dfrac{\partial\psi}{\partial
t}\right|_{x=a(t)}\dfrac{dt}{d\tau}=\sqrt{\dfrac{\gamma}{2k_BT(0)}}\dfrac{T(0)}{
T(t)}\dfrac{\kappa(t)}{\gamma}e^{-\Omega(t)}a(t).
\end{equation}
In general, this expression is not independent on time, except in the two cases
\begin{equation}
a(t)\propto\dfrac{T(t)}{\kappa(t)}e^{\Omega(t)}\quad\text{and}\quad a(t)=0.
\end{equation}
Interestingly, it is always possible to compute the FPT for the origin, which
is the point of symmetry of the potential. The FPTD for $a=0$ reads
\begin{eqnarray}
\nonumber
\wp(0,t|x_0)&=&\sqrt{\dfrac{\gamma}{2k_BT(0)}}\dfrac{T(t)}{T(0)}\dfrac{|x_0|
e^{2\Omega(t)}}{\sqrt{2\pi\left(\int_0^t\frac{T(s)}{T(0)}e^{2\Omega(s)}ds
\right)^3}}\exp\left(-\dfrac{\gamma}{2k_BT(0)}\dfrac{x_0^2}{2\int_0^t\frac{
T(s)}{T(0)}e^{2\Omega(s)}ds}\right).
\label{eqn:FPT_OU}
\end{eqnarray}
The exact FPTD for the Ornstein-Uhlenbeck process is available analytically
in literature only in the case of constant parameters \cite{siegert,
ricciardi1977} (or see \cite{grebenkov} for a review). Thus expression
\eqref{eqn:FPT_OU} also constitutes a novel result of this paper. 
Having at hand the explicit form of the FPTD can enhance the
quantitative study of feedback-control protocols, such as the ones in
\cite{garrahan_pre,saha_prl,tohme,archambault_epl,archambault_arxiv}. In these
references the authors study overdamped diffusions subjected to an
information-like control feedback, i.e., whenever the particle reaches a
specific threshold the stiffness of the potential is instantaneously set to
another value in order to maximise the work extraction. We illustrate results
\eqref{eqn:propagator_OU} and \eqref{eqn:FPT_OU} in the next subsection for
the isoentropic protocol.

\subsection{Isoentropic protocol}

\begin{figure}
\centering
\includegraphics[width=0.45\linewidth]{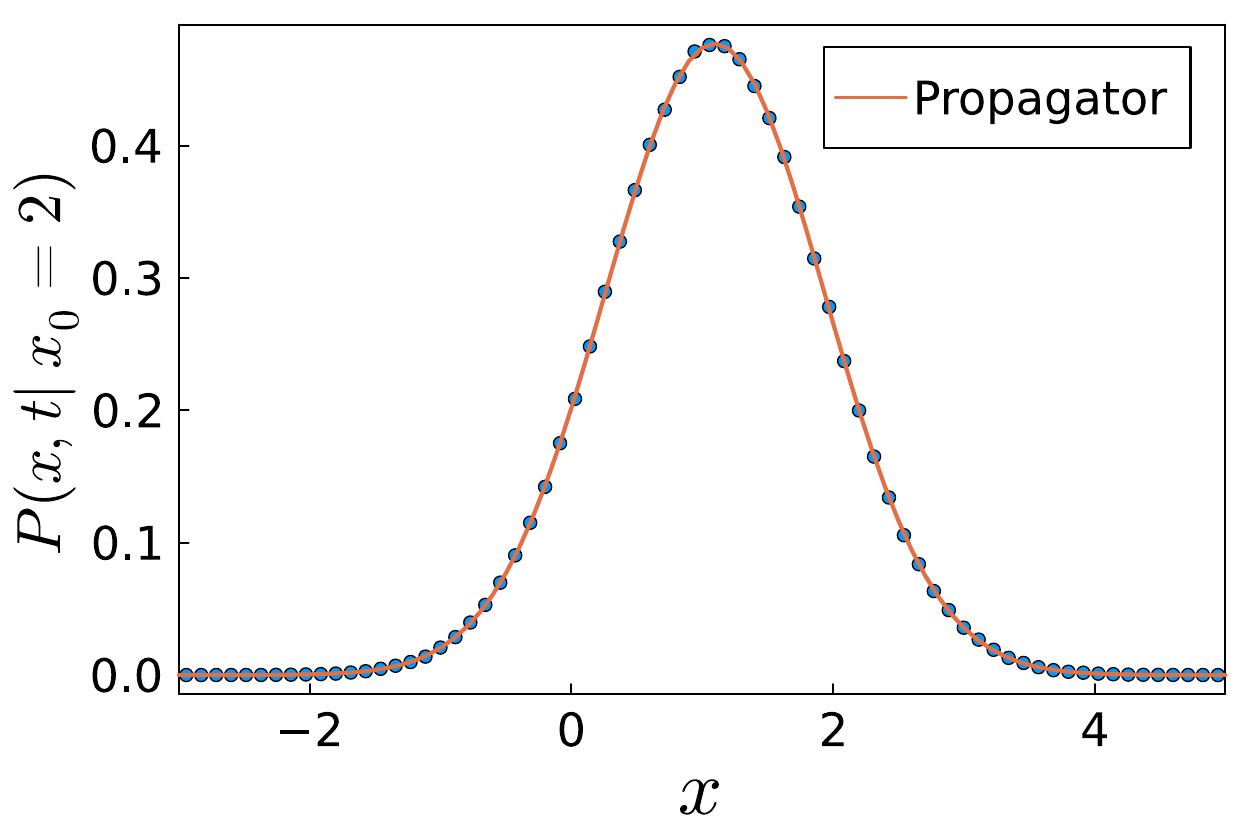}
\includegraphics[width=0.45\linewidth]{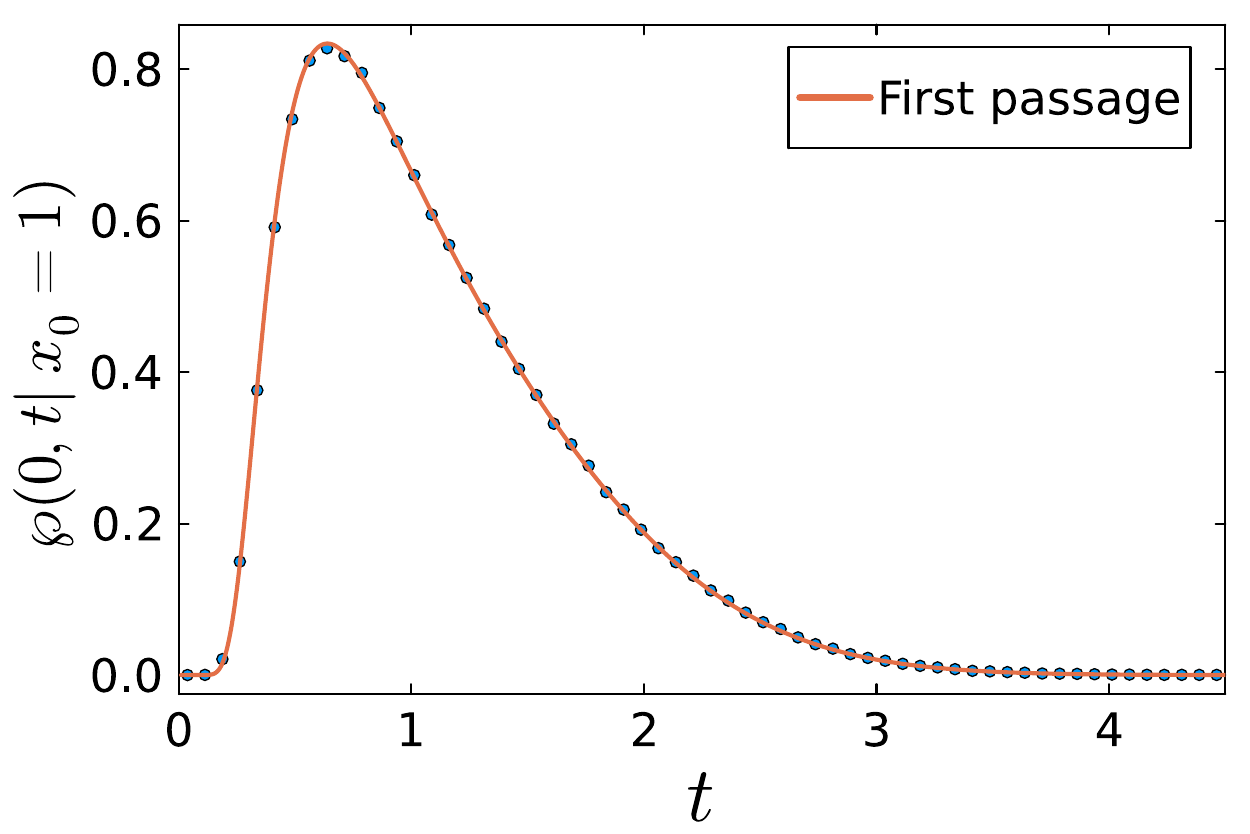}\\
(a) Propagator \hspace*{4.8cm} (b) FPTD
\caption{Comparison between analytical results (solid lines)
\eqref{eqn:propagator_isoentropic} for the propagator and
\eqref{eqn:FPT_isoentropic} for the FPTD versus simulations (dots). In both
cases the numerical simulations were based on the Euler-Maruyama scheme of
the Langevin equation \eqref{eqn:langevin}. In panel (a) we generated $2\times
10^6$ trajectories with parameters $x_0=2$, $\gamma=1$, $\kappa(t)=0.1+t$, and
$\kappa(0)/k_BT(0)=1$. In panel (b) we generated $10^6$ trajectories with the
parameters $x_0=1$, $\gamma=1$, $\kappa(t)=0.1+t$, and $\kappa(0)/k_B T(0)=1$.
Excellent agreement is observed.}
\label{fig:isoent}
\end{figure}

The so-called isoentropic (or pseudo-adiabatic) protocol
\cite{Schmiedl,martinez2015adiabatic} was introduced in the
field of stochastic thermodynamics as a building block of Carnot-type cycles
in colloidal heat engines \cite{martinez2017colloidal}. For overdamped
Langevin dynamics in a time-dependent harmonic potential and time-dependent
temperature, the protocol consists in having both temperature and stiffness
explicitly time-dependent while keeping their ratio constant. In mathematical
terms,
\begin{equation}
\dfrac{T(t)}{\kappa(t)}=\dfrac{T(0)}{\kappa(0)}.
\end{equation}
Such a protocol ensures that, if the initial condition is
equilibrium, the PDF associated with the particle position (and hence also
its Shannon entropy) is conserved in time. This is why such a protocol is
called adiabatic and isoentropic in the literature \cite{Schmiedl,martinez}.
Below we provide further insights into these features with our analytical
formalism. Imposing this relationship between $T(t)$ and $\kappa(t)$ we get
a simplified formula for $\tau(t)$,
\begin{equation}
\tau(t)=\dfrac{\gamma}{2\kappa(0)}\int_0^t2\Omega'(s)e^{2\Omega(s)}ds=\dfrac{
\gamma}{2\kappa(0)}\left(e^{2\Omega(t)}-1\right),
\end{equation}
hence the propagator \eqref{eqn:propagator} takes the simplified form
\begin{equation}
\label{eqn:propagator_isoentropic}
P(x,t|x_0,0)=\sqrt{\dfrac{\kappa(0)}{k_BT(0)}}\dfrac{1}{\sqrt{2\pi\left(1-
e^{-2\Omega(t)}\right)}}\exp\left(-\dfrac{\kappa(0)}{k_BT(0)}\dfrac{(x-x_0
e^{-\Omega(t)})^2}{2\left(1-e^{-2\Omega(t)}\right)}\right).
\end{equation}
The FPTD \eqref{eqn:FPT_distr} also has a simplified form, namely,
\begin{equation}
\label{eqn:FPT_isoentropic}
\wp(0,t|x_0)=\sqrt{\dfrac{\kappa(0)}{k_BT(0)}}\dfrac{2\kappa(t)}{\gamma}
\dfrac{|x_0|e^{2\Omega(t)}}{\sqrt{2\pi\left(e^{2\Omega(t)}-1\right)^3}}
\exp\left(-\dfrac{\kappa(0)}{k_BT(0)}\dfrac{x_0^2}{2\left(e^{2\Omega(t)}
-1\right)}\right).
\end{equation}
While the propagator \eqref{eqn:propagator_isoentropic} was known in
literature \cite{polyanin}, the FPT statistics, such as the FPTD
\eqref{eqn:FPT_isoentropic} are, to our knowledge, so far unknown for
adiabatic protocols. An excellent agreement between results
\eqref{eqn:propagator_isoentropic} and \eqref{eqn:FPT_isoentropic} and
numerical simulations based on realisations of the Langevin equation
\eqref{eqn:langevin} is demonstrated in figure \ref{fig:isoent}.

\begin{figure}
\centering
\includegraphics[width=0.48\linewidth]{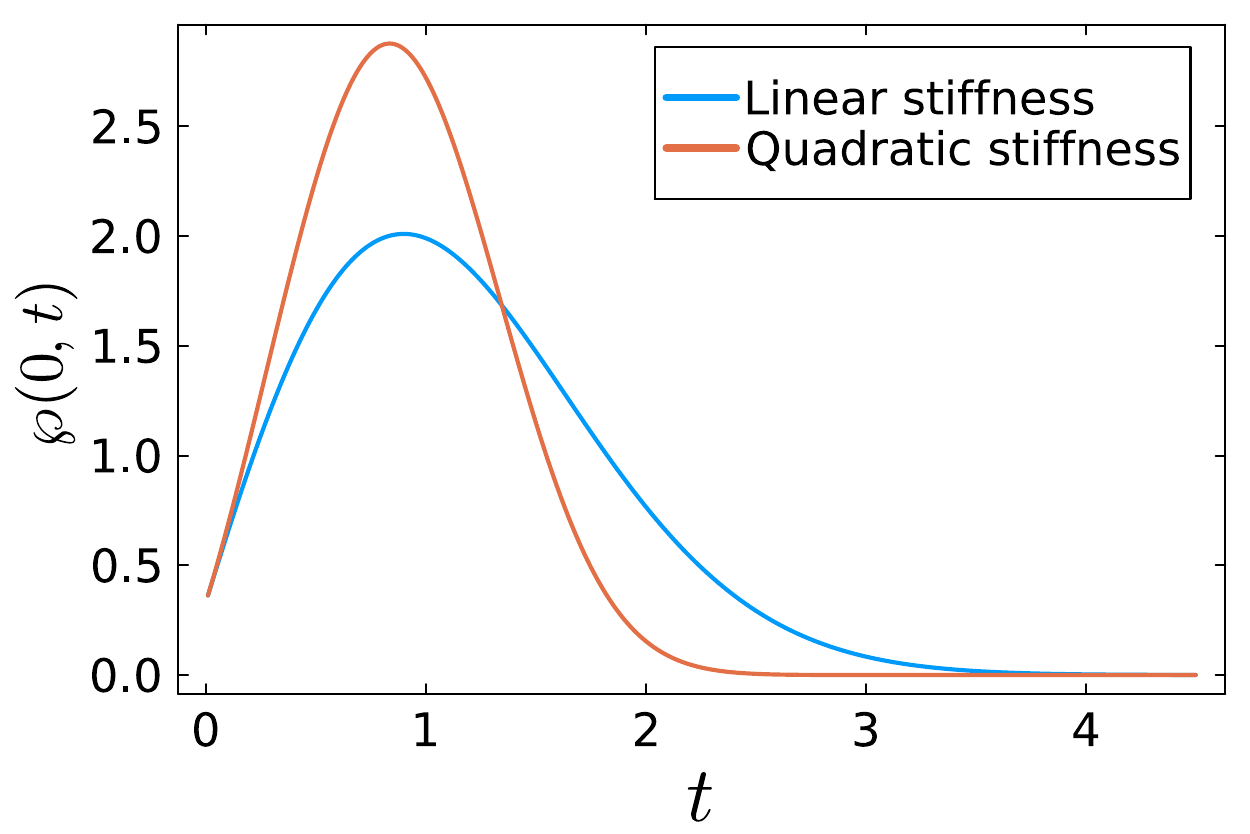}
\includegraphics[width=0.48\linewidth]{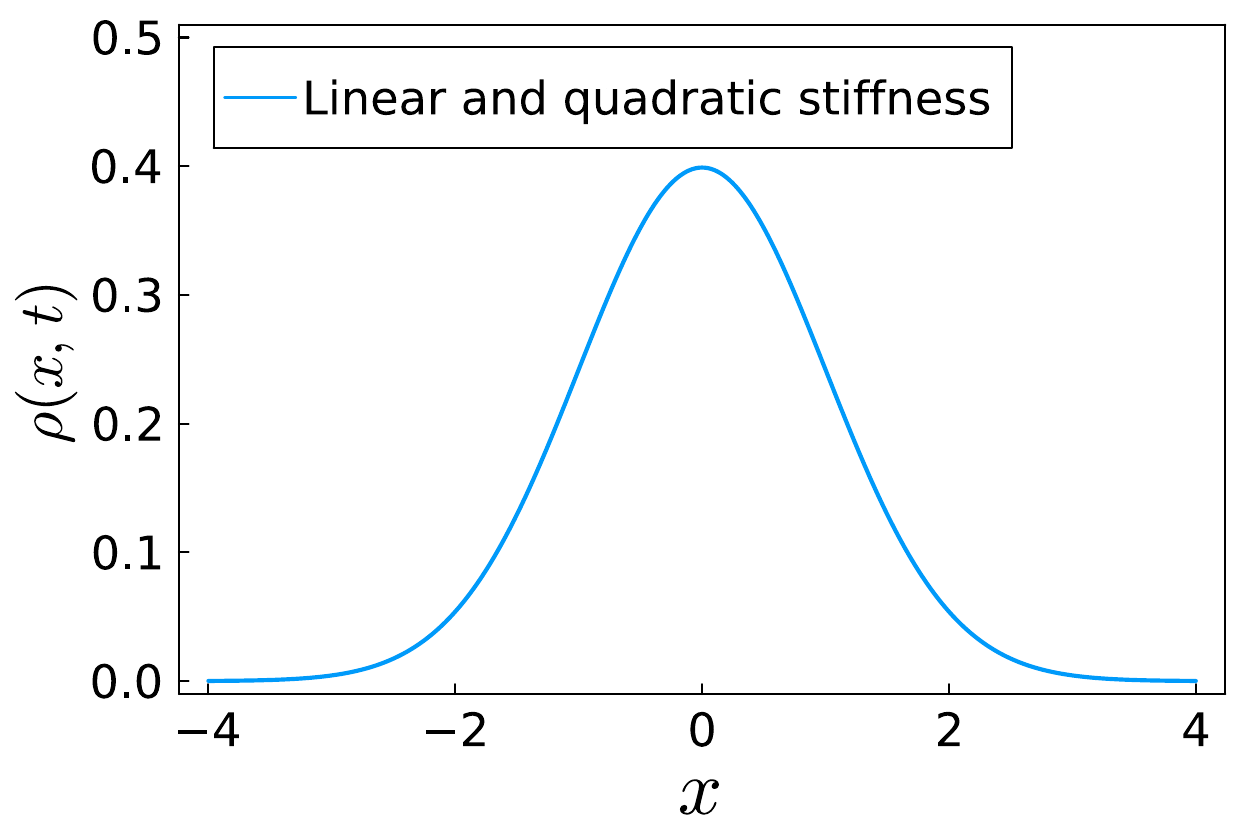}
(a) Averaged FPTD $\wp(0,t)$ \hspace*{3.6cm} (b) Density $\rho(x,t)$
\caption{(a) Comparison of the FPTD averaged over initial positions for two
different choices of the stiffness protocol $\kappa(t)$. For the blue curve
we used the linear function $\kappa(t)=T(t)=0.1+t$, while for the orange
curve we used the quadratic form $\kappa(t)=T(t)=0.1+t+t^2$. In both cases
we use $\gamma=1$. (b) Density $\rho(x,t)$, which is equal for the two cases
of linear and quadratic stiffness.}
\label{fig:averaged_FPT}
\end{figure}

We conclude this subsection with an observation concerning the average over
initial positions. Clearly, if the system starts at equilibrium, the form of
the probability distribution is preserved over time. More explicitly, if we
denote with $\rho(x,t)$ the solution of the Fokker-Planck equation
\eqref{eqn:FPE} with initial condition
\begin{equation}
\rho(x,0)=\sqrt{\dfrac{\kappa(0)}{k_BT(0)}}\dfrac{1}{\sqrt{2\pi}}\exp\left(
-\dfrac{\kappa(0)}{k_BT(0)}\dfrac{x^2}{2}\right),
\end{equation}
it is possible to obtain $\rho(x,t)$ via a convolution of the initial
condition with the propagator \eqref{eqn:propagator_isoentropic},
\begin{equation}
\rho(x,t)=\int_{-\infty}^{\infty}\rho(x_0,0)P(x,t|x_0)dx_0,
\end{equation}
which clearly shows that $\rho(x,t)=\rho(x,0)$, hence the form of the
distribution does not change with time for any choice of the protocols
$\kappa(t)$ and $T(t)$. Nevertheless, the FPTD averaged over initial
positions is not protocol-independent and reads
\begin{equation}
\wp(0,t)\equiv\int_{-\infty}^{+\infty}\rho(x_0,0)\wp(0,t|x_0)dx_0=\dfrac{
2}{\pi}\dfrac{\kappa(t)}{\gamma}\dfrac{1}{\sqrt{e^{2\Omega(t)}-1}}.
\end{equation}
This is highlighted in figure \ref{fig:averaged_FPT}. This example shows
that knowing only the distribution of the process may not be enough to
infer any specific property of the FPTD.

\subsection{Stochastic Gompertz model}

An important process for population dynamics is the
stochastic Gompertz model proposed in~\cite{capocelli1974} and studied
further in~\cite{skiadas,ricciardi1999}. The SDE describing the process reads
\begin{equation}
\label{eqn:stoch_gompertz}
dX_t=-\alpha(t)X_t\ln\left(\dfrac{X_t}{K(t)}\right)dt+\beta(t)X_tdW_t.
\end{equation}
The deterministic version of the model (without noise) was introduced two
centuries ago \cite{gompertz} to express the law of human mortality,
and it was more recently applied to the modelling of
cellular growth of tumours \cite{laird}. In the
existing literature \cite{skiadas,capocelli1974,ricciardi1999} the parameters
$\alpha$, $K$, and $\beta$ were considered as constants. Here, instead,
with the help of our formalism, we consider them as arbitrary functions of
time. Note that if $\alpha(t)>0$, the drift term in~\eqref{eqn:stoch_gompertz}
is positive for $X_t<K(t)$ and negative for $X_t>K(t)$, while it vanishes for
$X_t=K(t)$. Therefore, $K(t)$ represents a carrying capacity, which might
depend on time if the resources of the environment are not constant, while
$\alpha(t)$ represents either a birth or a mortality rate. The stochastic
motion starts at some $X_0=x_0>0$ and it is confined to be positive at
all times.

\begin{figure}
\centering
\includegraphics[width=0.48\linewidth]{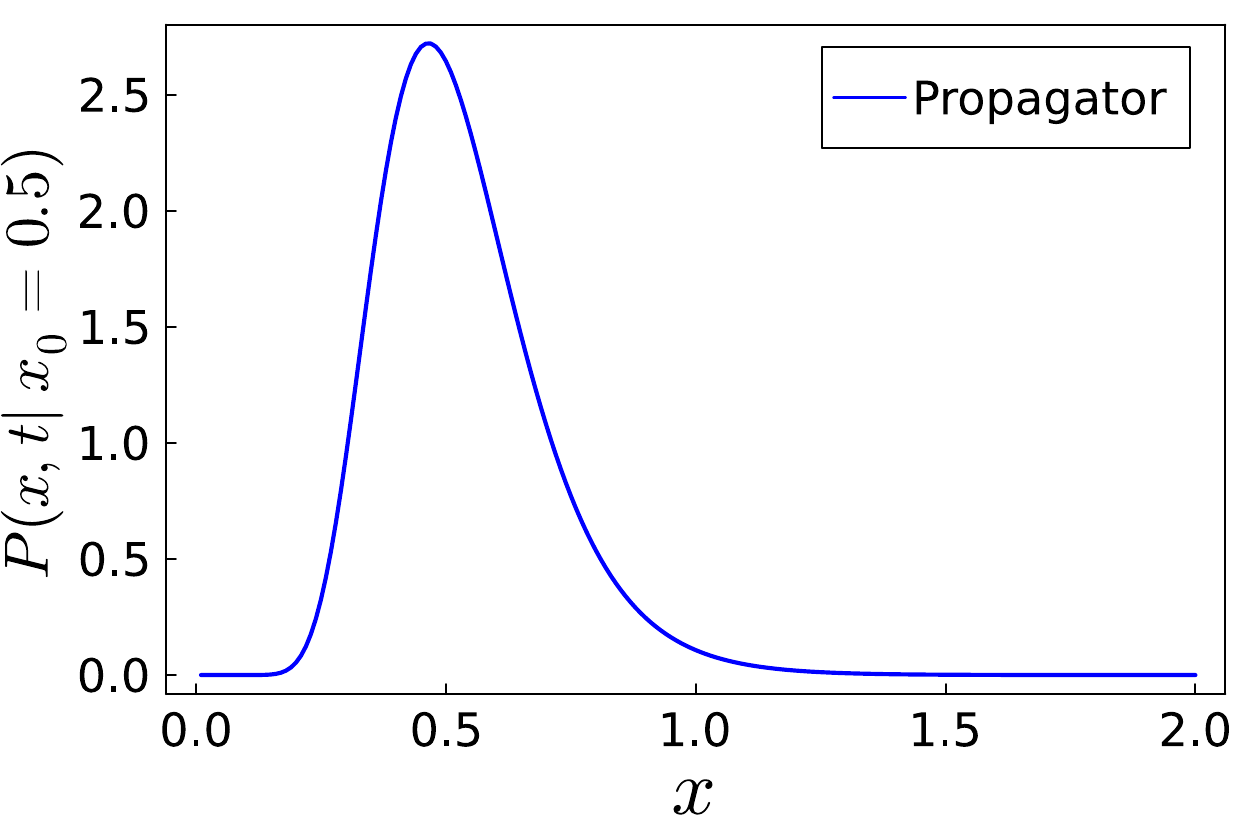}
\includegraphics[width=0.48\linewidth]{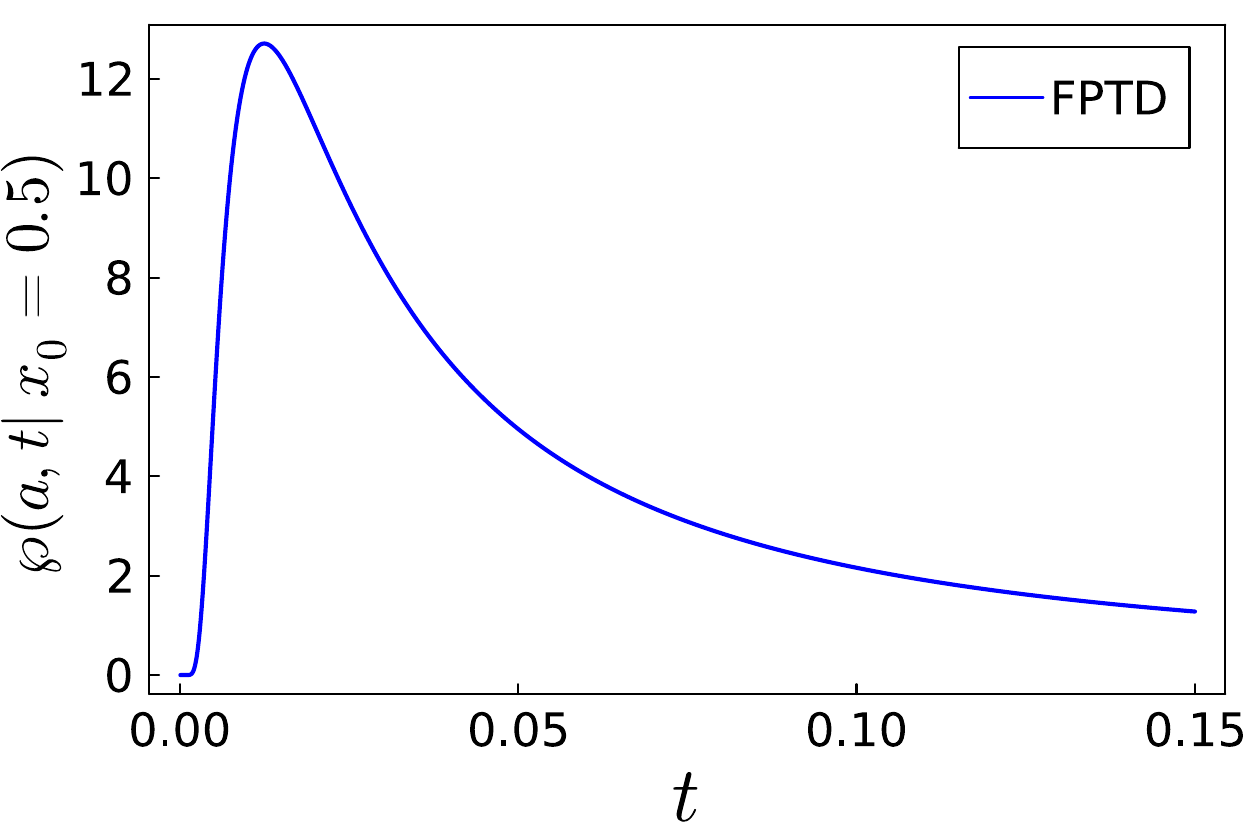}
(a) Propagator $P(x,t|x_0)$ \hspace*{3.0cm} (b) FPTD $\wp\left(K\exp\left(
-\frac{\beta^2}{2\alpha}\right)t|x_0\right)$
\caption{(a) Propagator \eqref{eqn:gompertz_propagator} and (b)
FPTD \eqref{eqn:gompertz_FPTD} for the stochastic Gompertz model
\eqref{eqn:stoch_gompertz} with constant parameters. The parameters are
$\alpha=\beta=K=1$ and $x_0=0.5$. The propagator is plotted for $t=0.1$.}
\label{fig:gompertz}
\end{figure}

Let us now have a look at the form of the propagator. First, using equation
\eqref{eqn:cherkasov} the Cherkasov function reads
\begin{equation}
\mathcal{C}(t)=\dfrac{d\ln\beta(t)}{dt}+\alpha(t),
\end{equation}
which, being independent of $x$, allows us to write the propagator without
explicitly solving the Fokker-Planck equation, thus avoiding the calculations
in \cite{capocelli1974}. The transforming function $\psi(x,t)$ can be obtained
via relation \eqref{eqn:psi}, producing
\begin{eqnarray}
\nonumber
\psi(x,t|x_0)&=&\dfrac{1}{\beta(0)}\exp\left(\int_0^t\alpha(t')dt'\right)\ln
\left(\dfrac{x}{x_0}\right)\\
&&+\int_0^t\left(\dfrac{\beta(t')}{2}+\dfrac{\alpha(t')\ln\left(\frac{x_0}
{K(t')}\right)}{\beta(t')} \right)\dfrac{\beta(t')}{\beta(0)}\exp\left(\int_0
^{t'}\alpha(s)ds\right)dt';
\end{eqnarray}
while $\tau(t)$, using~\eqref{eqn:tau}, reads
\begin{equation}
\tau(t)=\dfrac{1}{\beta^2(0)}\int_0^t\beta^2(t')\exp\left(2\int_0^{t'}\alpha(s)ds
\right)dt'.
\end{equation}
Here we set $t_0=0$ everywhere. The propagator is then obtained from plugging
the forms of $\psi$ and $\tau$ into equation \eqref{eqn:propagator}. In the
simplest case of constant parameters the propagator then reads
\begin{equation}
\label{eqn:gompertz_propagator}
P(x,t|x_0)=\dfrac{1}{x\sqrt{2\pi\frac{\beta^2}{2\alpha}\left(1-e^{2\alpha t}
\right)}}\exp\left(-\dfrac{\left(\ln\left(\frac{x}{K}\right)+\frac{\beta^2}{
2\alpha}\left(1-e^{-\alpha t}\right)-\ln\left(\frac{x_0}{K}\right)e^{-\alpha
t}\right)^2}{\frac{\beta^2}{2\alpha}\left(1-e^{2\alpha t}\right)}\right),
\end{equation}
which is plotted in Fig.~\ref{fig:gompertz}a.
The stationary distribution can then be easily obtained as
\begin{equation}
P^{(s)}(x)=\dfrac{1}{x\sqrt{2\pi\frac{\beta^2}{2\alpha}}}\exp\left(-\dfrac{
\left(\ln\left(\frac{x}{K}\right)+\frac{\beta^2}{2\alpha}\right)^2}{\frac{
\beta^2}{2\alpha}}\right).
\end{equation}

For the FPTD, we need to check when condition \eqref{eqn:condition2} is
satisfied. We compute first the derivative
\begin{equation}
\left.\dfrac{d\psi}{d\tau}\right|_{x=a(t)}=\beta(0)\left[\dfrac{\alpha(t)}{
\beta^2(t)}\ln\left(\dfrac{a(t)}{K(t)}\right)+\dfrac{1}{2}\right]\exp\left(
-\int_0^t\alpha(t')dt'\right)
\end{equation}
and then require that it must be independent of $t$. A possibility is thus
when $a(t)$ has the form
\begin{equation}
a(t)=K(t)\exp\left(-\frac{\beta^2(t)}{2\alpha(t)}\right).
\end{equation}
For this specific choice of the boundary, considering again the case of
constant parameters, the FPTD reads
\begin{equation}
\label{eqn:gompertz_FPTD}
\wp\left(K\exp\left(-\frac{\beta^2}{2\alpha}\right),t|x_0\right)=\dfrac{
\left|\frac{\beta^2}{2\alpha}+\ln\left(\frac{x_0}{K}\right)\right|}{\sqrt{2
\pi\frac{\beta^2}{(2\alpha)^3}\left(e^{2\alpha t}-1\right)^3}}\exp\left(2
\alpha t-\dfrac{\left(\frac{\beta^2}{2\alpha}+\ln\left(\frac{x_0}{K}\right)
\right)^2}{2\frac{\beta^2}{2\alpha}\left(e^{2\alpha t}-1\right)}\right),
\end{equation}
which is displayed in Fig.~\ref{fig:gompertz}b.

\section{Connections with existing results}
\label{exist}

As we already discussed, our formalism can provide solutions to many non-linear
SDEs. We here illustrate a class of such models satisfying condition
\eqref{eqn:condition}. A particular choice of the coefficients $\mu$ and
$\sigma$ satisfying condition \eqref{eqn:condition} is the product form
\begin{equation}
\label{eqn:separability_sigma}
\sigma(x,t)\equiv\sigma_x(x)\sigma_t(t)
\end{equation}
i.e., $\sigma$ can be separated into two functions, the purely $x$-dependent
$\sigma_x(x)$ and the purely $t$-dependent $\sigma_t(t)$. Then, for $\mu$ we
have
\begin{equation}
\label{eqn:relation_mu_sigma}
\mu(x,t)=\dfrac{1}{2}\dfrac{\partial}{\partial x}\left(\dfrac{\sigma^2(x,t)}{2}
\right).
\end{equation}
Interestingly, it can be shown that for this choice of the parameters, the
SDE can be rewritten in the Stratonovich formalism as $dX_t=\sigma_t(t)\sigma
_x(X_t)\circ dW_t$.
The two equations \eqref{eqn:separability_sigma} and \eqref{eqn:relation_mu_sigma}
imply that both conditions \eqref{eqn:condition} and \eqref{eqn:condition2} hold,
in particular \eqref{eqn:condition2} holds for any boundary $a(t)$. Indeed, using
expressions \eqref{eqn:separability_sigma} and \eqref{eqn:relation_mu_sigma}, we
get that $\psi$ and $\tau$ take the simplified expressions
\begin{equation}
\psi(x,t|x_0,t_0)=\dfrac{1}{\sigma_t(t_0)}\int_{x_0}^x\dfrac{dx'}{\sigma_x(x')}
\end{equation}
and
\begin{equation}
\tau(t|t_0)=\dfrac{1}{\sigma_t^2(t_0)}\int_{t_0}^t\sigma_t^2(s)ds.
\end{equation}
Notice that $\psi$ does not depend on time here, therefore condition
\eqref{eqn:condition2} holds for any boundary $a(t)$. This prompts the
following observation: for all physical systems in which $\mu(x,t)$ is
a potential force, i.e., $\mu(x,t)=-\dfrac{1}{\gamma}\dfrac{\partial}{\partial x}U(x,t)$,
and $\sigma^2$ is related to the temperature as $\sigma^2(x,t)=2k_BT(x,t)/
\gamma$, equation \eqref{eqn:relation_mu_sigma} implies that
\begin{equation}
\label{eqn:U_over_T}
\dfrac{U(x,t)}{k_BT(x,t)}=\mathrm{const}.
\end{equation}
Equation~\eqref{eqn:U_over_T} implies that the process is purely diffusive, indeed, as stated before, the SDE reads $dX_t=\sigma(x,t)\circ dW_t$, thus the stationary distribution of the process is independent of time and space for the case of closed or periodic boundary conditions. Notably,
our results show the FPTD can be known analytically
for rather generic boundary conditions for all processes for which \eqref{eqn:U_over_T} holds.
Such processes were thoroughly studied  by e.g. H\"anggi, Talkner and Borkovec  in Ref. \cite{hanggi_review} where they outline explicit results to obtain all
moments of the FPT for processes with $U/k_BT=\mathrm{const}$, yet the full
FPTD has not been investigated. Our formalism allows the determination of
this entire FPTD. Moreover, processes with $U/k_BT=\mathrm{const}$ are a
subclass of the wider class described by equations \eqref{eqn:condition} and
\eqref{eqn:condition2}.

\section{Conclusions}
\label{concl}

We discussed a method to obtain exact solutions of a class of
time-inhomogeneous stochastic differential equations and their associated
FPT problems.  This method represents a valid and simpler
alternative to other approaches. The technique discussed in this paper for
FPTDs, while mentioned in~\cite{ricciardi1984}, to the best of
our knowledge, has been unused and largely
unknown to the statistical physics community for a long time. The explicit
form of FPTDs can highly enhance the quantitative study of feedback-control
protocols, such as the ones in \cite{garrahan_pre, saha_prl, tohme,
archambault_epl, archambault_arxiv},
and can provide useful insights in experiments with optical tweezers beyond those exemplified in Sec.~\ref{sec:OU}. 
To this end we identified a class of
one-dimensional problems with time-dependent parameters that are amenable
to exact solutions, yet extensions to higher dimensions and more complex
scenarios (e.g., with nonlinear time dependencies in the boundaries) are
possible and will be the topic of future work.  Moreover,
having at hand analytical results for multidimensional models could
provide interesting insights into non-Markovian processes via Markovian
embedding~\cite{loos}. With our method, we were able to
generalise well-known results and to obtain new solutions to non-autonomous
problems. We were able to find the propagators
\eqref{eqn:propagator_abs} and \eqref{eqn:propagator_ref} in the constrained
domain that could be useful for further studies in constrained random
walks~\cite{de2021generating}. We only considered models with deterministic
parameters, but further generalisations to models such as diffusing diffusivity
\cite{chubynsky_DD,jain,Sposini_2020} or the so-called $\textrm{OU}^2$
process \cite{cocconi_ou2} should be rather straightforward. Within
statistical physics, we believe that our results could find prominent
applications, inter alia, within control theory \cite{bechhoefer2021control},
non-equilibrium calorimetry \cite{khodabandehlou2024vanishing}, or computing
\cite{kolchinsky2021work}.  We believe that our work will serve as a
reference also beyond the statistical physics community in other areas,
in which stochastic processes and FPTs are important.

\ack

We acknowledge partial funding by the German Research Foundation (DFG,
grant nos. ME 1535/13-1 and 1535/16-1).

\appendix

\section{Cherkasov condition and explicit form of transforming functions}
\label{appendix1}

Here we will prove equations \eqref{eqn:condition}, \eqref{eqn:SDE_solution},
\eqref{eqn:propagator}, \eqref{eqn:psi}, and \eqref{eqn:tau}. The dynamic
of the position $X_t$ is governed by the SDE (\ref{eqn:Ito_SDE}), which
we here repeat for the convenience of the reader:
\begin{equation}
dX_t=\mu(X_t,t)dt+\sigma(X_t,t)dW_t,
\end{equation}
where $W_t$ is the standard Wiener process. Consider now the transformation
$Y_t=\psi(X_t,t)$; according to the It\^{o} rule, the SDE for $Y_t$ reads
\begin{equation}
dY_t=\left(\dfrac{\partial\psi}{\partial t}\biggr|_{x=X_t}+\mu(X_t,t)\dfrac{
\partial\psi}{\partial x}\biggr|_{x=X_t}+\dfrac{\sigma^2(X_t,t)}{2}\dfrac{
\partial^2\psi}{\partial x^2}\biggr|_{x=X_t}\right)dt+\sigma(X_t,t)\dfrac{
\partial\psi}{\partial x}\biggr|_{x=X_t}dW_t.
\end{equation}
In order to be mappable onto the Wiener process, both drift and volatility
of the new process should be independent of the position. Therefore
\begin{equation}
\label{eqn:system}
\begin{cases}
\dfrac{\partial\psi(x,t)}{\partial t}+\mu(x,t)\dfrac{\partial\psi(x,t)}{
\partial x}+\dfrac{\sigma^2(x,t)}{2}\dfrac{\partial^2\psi(x,t)}{\partial
x^2}=\mu_Y(t),\\[0.2cm]
\sigma(x,t)\dfrac{\partial\psi(x,t)}{\partial x}=\sigma_Y(t),
\end{cases}
\end{equation}
where $\mu_Y(t)$ and $\sigma_Y(t)$ are, respectively, the new drift and
variance, both solely time-dependent. We anticipate that in the derivation
below it will be possible to set $\mu_Y(t)=0$. From equation \eqref{eqn:system}
we obtain 
\begin{equation}
\label{eqn:dpsi_dx}
\dfrac{\partial\psi(x,t)}{\partial x}=\dfrac{\sigma_Y(t)}{\sigma(x,t)}>0,
\end{equation}
which proves that $\psi(x,t)$ is strictly increasing in the first variable $x$.
Substituting this result into the first equation of \eqref{eqn:system} we get
\begin{equation}
\dfrac{\partial\psi(x,t)}{\partial t}=\sigma_Y(t)\left(\dfrac{1}{2}\dfrac{
\partial\sigma(x,t)}{\partial x}-\dfrac{\mu(x,t)}{\sigma(x,t)}\right)+\mu_Y(t).
\end{equation}
If the second partial derivatives of $\psi$ are continuous, we know from
Schwarz's theorem
that $\partial^2\psi/\partial x\partial t=\partial^2\psi/\partial t\partial
x$, and therefore
\begin{equation}
\dfrac{d\sigma_Y}{dt}\dfrac{1}{\sigma}-\dfrac{\sigma_Y}{\sigma^2}\dfrac{
\partial\sigma}{\partial t}=\sigma_Y\dfrac{\partial}{\partial x} \left(
\dfrac{1}{2}\dfrac{\partial\sigma}{\partial x}-\dfrac{\mu}{\sigma}\right).
\end{equation}
Since there is no risk of confusion, we here drop the arguments of $\sigma$,
$\sigma_Y$, and $\mu$. From some simple manipulations we get
\begin{equation}
\label{eqn:cherkasov_appendix}
\dfrac{1}{\sigma_Y(t)}\dfrac{d\sigma_Y(t)}{dt}=\dfrac{1}{\sigma}\dfrac{\partial
\sigma}{\partial t}+\sigma\dfrac{\partial}{\partial x}\left(\dfrac{1}{2}\dfrac{
\partial\sigma}{\partial x}-\dfrac{\mu}{\sigma}\right)\equiv\mathcal{C}(t),
\end{equation}
where the left hand side does not depend on $x$---thus by taking the
partial derivative $\partial/\partial x$ we immediately see that the
condition \eqref{eqn:condition} on the Cherkasov function is indeed
fulfilled. In other words, if the drift and the volatility of the new
process are space-independent then the Cherkasov function fulfils condition
\eqref{eqn:condition}, and the inverse is also true.  This formally proves
that condition \eqref{eqn:condition} is necessary and sufficient.

Moreover, by solving the previous differential equation, we get the form
\begin{equation}
\label{eqn:sigma_prime}
\sigma_Y(t)=\exp\left(\int_{t_0}^t\mathcal{C}(s)ds\right)
\end{equation}
of $\sigma_Y(t)$, where, since it is arbitrary, for simplicity we set $\sigma
_Y(t_0)=1$. Substituting \eqref{eqn:sigma_prime} into \eqref{eqn:dpsi_dx} and
integrating over $x$ we obtain the explicit formula of the transformation
\begin{equation}
\label{eqn:psi_psi0}
\psi(x,t)=\exp\left(\int_{t_0}^t\mathcal{C}(s)ds\right)\int_{x_0}^x\dfrac{1}{
\sigma(x',t)}dx'+\psi(x_0,t).
\end{equation}
So far we have not specified the boundary condition $\psi(x_0,t)$, which is
arbitrary. 

Let us show that $\psi(x_0,t)$ and $\mu_Y(t)$ are related one-to-one. To see
this we substitute expression \eqref{eqn:psi_psi0} back into the first
equation of the system \eqref{eqn:system},
\begin{eqnarray}
\nonumber
\mu_Y(t)&=&\dfrac{d\psi(x_0,t)}{dt}+\left[\mathcal{C}(t)\int_{x_0}^x\dfrac{
1}{\sigma(x',t)}dx'-\int_{x_0}^x\dfrac{1}{\sigma^2(x',t)}\dfrac{\partial
\sigma}{\partial t}dx'\right.\\
&&\left.+\dfrac{\mu(x,t)}{\sigma(x,t)}-\dfrac{1}{2}\dfrac{\partial\sigma}{
\partial x}\right]\exp\left(\int_{t_0}^t\mathcal{C}(t')dt'\right).
\label{muy}
\end{eqnarray}
We notice that the term appearing on the right hand side of this equation can
be simplified by use of the definition of the Cherkasov function
\eqref{eqn:cherkasov},
\begin{equation}
\int_{x_0}^x\dfrac{1}{\sigma^2(x',t)}\dfrac{\partial\sigma}{\partial t}dx'+
\dfrac{1}{2}\dfrac{\partial\sigma}{\partial x}-\dfrac{\mu(x,t)}{\sigma(x,t)}
=\mathcal{C}(t)\int_{x_0}^x\dfrac{1}{\sigma(x',t)}dx'-\dfrac{\mu(x_0,t)}{
\sigma(x_0,t)}+\dfrac{1}{2}\dfrac{\partial\sigma}{\partial x}\biggr|_{(x_0,t)},
\end{equation}
which simplifies equation \eqref{muy} for $\mu_Y(t)$,
\begin{equation}
\mu_Y(t)=\dfrac{d\psi(x_0,t)}{dt}+\left( \dfrac{\mu(x_0,t)}{\sigma(x_0,t)}-
\dfrac{1}{2}\dfrac{\partial\sigma}{\partial x}\biggr|_{(x_0,t)}\right)\exp
\left(\int_{t_0}^t\mathcal{C}(t')dt'\right).
\end{equation}
Thus we found a differential equation for $\psi(x_0,t)$. As $\psi(x_0,t)$ is
arbitrary, a convenient choice is the one leading to $\mu_Y(t)=0$, yielding
\begin{equation}
\psi(x_0,t)=\int_{t_0}^tds\left(\dfrac{1}{2}\dfrac{\partial\sigma}{\partial x}
\biggr|_{(x_0,s)}-\dfrac{\mu(x_0,s)}{\sigma(x_0,s)}\right)\exp\left(\int_{t_0}
^s\mathcal{C}(s')ds'\right),
\end{equation}
where we set $\psi(x_0,t_0)=0$. Therefore, the final form of $\psi(x,t)$ is
\begin{eqnarray}
\nonumber
\psi(x,t)&=&\exp\left(\int_{t_0}^t\mathcal{C}(s)ds\right)\int_{x_0}^x\dfrac{
1}{\sigma(x',t)}dx'\\
&&+\int_{t_0}^tds\left(\dfrac{1}{2}\dfrac{\partial\sigma}{\partial x}\biggr|_{
(x_0,s)}-\dfrac{\mu(x_0,s)}{\sigma(x_0,s)}\right)\exp\left(\int_{t_0}^s
\mathcal{C}(s')ds'\right).
\end{eqnarray}
This transformation maps the SDE onto $dY_t=\sigma_Y(t)dW_t$ and can indeed
be considered as a modification of the Lamperti transform \cite{lamperti}.

Finally, the function reindexing time reads
\begin{equation}
\label{eqn:tau_appendix}
\tau(t)=\int_{t_0}^t\sigma_Y^{2}(s)ds=\int_{t_0}^t\exp\left(2\int_{t_0}^s
\mathcal{C}(r)dr\right)ds,
\end{equation}
as we set out to prove. We conclude with a final remark on $\psi(x,t)$ and
$\tau(t)$. Namely, there are infinitely many transformations serving our
purpose; for simplicity, we chose the specific forms \eqref{eqn:psi} and
\eqref{eqn:tau}, that vanish at the initial points $x_0,t_0$.

\section{Extension to inhomogeneous geometric Brownian motion}
\label{app:second_cherkasov}

As stated in the main text, condition~\eqref{eqn:condition}
does not cover all SDEs that are solvable analytically. One prominent example
is the inhomogeneous geometric Brownian motion (IGBM) which is described by
the SDE
\begin{equation}
dX_t=\left(\alpha X_t+\beta(t)\right)dt+\sigma(t)X_tdW_t,
\end{equation}
where the parameters $\alpha$, $\beta$, and $\sigma$ are generic functions
of time. A complete solution to~\eqref{eqn:IGBM} is contained in the book
of Mao~\cite{Mao_book}.  One may argue that there must exist a {\em second
Cherkasov condition} that whenever satisfied, it is guaranteed that the SDE
is mappable onto IGBM.  We now derive such a condition.

Following a procedure identical to the one described in \ref{appendix1},
starting again from $dX_t = \mu(X_t, t) dt + \sigma(X_t, t) dW_t$ and applying
a generic transformation $Y_t = \psi(X_t,t)$ we end up at a new SDE for $Y_t$
with new drift $\mu_Y$ and a new volatility $\sigma_Y$. We suppose that the
transforming function $\psi(x,t)$ is sufficiently smooth, so that we can
take derivatives without caring about discontinuities. We now require the
forms $\mu_Y(y,t)=\alpha(t)y+\beta(t)$ and $\sigma_Y(y,t)=\zeta(t)y$. With
these requirements, the system~\eqref{eqn:system} becomes
\begin{equation}
\label{eqn:system2}
\begin{cases}
\dfrac{\partial\psi(x,t)}{\partial t}+\mu(x,t)\dfrac{\partial\psi(x,t)}{
\partial x}+\dfrac{\sigma^2(x,t)}{2}\dfrac{\partial^2\psi(x,t)}{\partial
x^2}=\alpha(t) \psi(x,t) + \beta(t),\\[0.2cm]
\sigma(x,t)\dfrac{\partial\psi(x,t)}{\partial x}=\zeta(t) \psi(x,t),
\end{cases}
\end{equation}
where we substituted $y$ with $\psi(x,t)$. From the second equation of
\eqref{eqn:system2} we get
\begin{equation}
\label{eqn:dpsi_dx2}
\dfrac{\partial\psi(x,t)}{\partial x}=\zeta(t)\dfrac{\psi(x,t)}{\sigma(x,t)},
\end{equation}
which substituted into the first equation, yields
\begin{equation}
\dfrac{\partial\psi}{\partial t}+\left[\dfrac{\mu}{\sigma}-\dfrac{1}{2}\dfrac{
\partial\sigma}{\partial x}+\dfrac{\zeta}{2}\right]\zeta\psi=\alpha\psi+\beta.
\end{equation}
Here, for convenience, we dropped the explicit dependence on $x,t$ everywhere.
Next, we divide by $\psi$ and we take the derivative with respect to $x$,
\begin{equation}
\dfrac{\partial}{\partial t}\left(\dfrac{\zeta}{\sigma}\right)+\dfrac{\partial}{
\partial x}\left(\dfrac{\mu}{\sigma}-\dfrac{1}{2}\dfrac{\partial\sigma}{\partial
x}\right)\zeta=\beta\dfrac{\partial}{\partial x}\left(\dfrac{1}{\psi}\right),
\end{equation}
where we used the fact that $\frac{\partial}{\partial x}\left(\frac{1}{\psi}
\frac{\partial\psi}{\partial t}\right)=\frac{\partial}{\partial t}\left(\frac{
1}{\psi}\frac{\partial\psi}{\partial x}\right)$ and that $\frac{\partial\zeta}{
\partial x}=\frac{\partial\alpha}{\partial x}=\frac{\partial\beta}{\partial x}=0$.
After further manipulations we arrive at
\begin{equation}
\dfrac{1}{\sigma}\dfrac{\partial\zeta}{\partial t}-\dfrac{\zeta}{\sigma^2}
\dfrac{\partial\sigma}{\partial t}+\dfrac{\partial}{\partial x}\left(\dfrac{
\mu}{\sigma}-\dfrac{1}{2}\dfrac{\partial\sigma}{\partial x}\right)\zeta=-
\dfrac{\beta}{\psi}\dfrac{\zeta}{\sigma}.
\end{equation}
We now multiply by $\sigma$ and divide by $\zeta$,
\begin{equation}
\dfrac{1}{\zeta}\dfrac{\partial\zeta}{\partial t}-\dfrac{1}{\sigma}\dfrac{
\partial\sigma}{\partial t}+\sigma\dfrac{\partial}{\partial x}\left(\dfrac{
\mu}{\sigma}-\dfrac{1}{2}\dfrac{\partial\sigma}{\partial x}\right)=-\dfrac{
\beta}{\psi}.
\end{equation}
On the left hand side we recognise the Cherkasov function \eqref{eqn:cherkasov}.
Thus, the previous equation can be rewritten as
\begin{equation}
\label{eqn:for_beta}
\dfrac{\partial\ln\zeta}{\partial t}=\mathcal{C}-\dfrac{\beta}{\psi}.
\end{equation}
We note that here, according to our assumption that the Cherkasov condition
\eqref{eqn:condition} does not hold, the function $\mathcal{C}(x,t)$ will
depend both on $x$ and $t$. Hence, taking a derivative with respect to $x$
the left hand side vanishes and we get
\begin{equation}
\dfrac{\partial}{\partial x}\left(\mathcal{C}-\dfrac{\beta}{\psi}\right)=0.
\end{equation}
The previous expression appears like another Cherkasov condition, yet there
is a dependence on $\beta$, which is an unknown parameter. We now reexpress
it in terms of the original parameters $\mu$ and $\sigma$ of the SDE. Using
again the expression for $\partial\psi/\partial x$ we get
\begin{equation}
\dfrac{\partial\mathcal{C}}{\partial x}+\dfrac{\beta}{\psi}\dfrac{\zeta}{
\sigma}=0.
\end{equation}
Further manipulating the previous expression, recalling that, by our
assumption, both $\beta$ and $\zeta$ do not depend on $x$, we get
\begin{equation}
\dfrac{\partial}{\partial x}\left(\sigma \dfrac{\partial\mathcal{C}}{\partial
x}\right)+\zeta\dfrac{\partial\mathcal{C}}{\partial x}=0,
\end{equation}
from which we find the explicit form of the function $\zeta(t)$,
\begin{equation}
\zeta(t)=-\left(\dfrac{\partial\mathcal{C}}{\partial x}\right)^{-1}\dfrac{
\partial}{\partial x}\left(\sigma\dfrac{\partial\mathcal{C}}{\partial x}\right),
\end{equation}
and also the second Cherkasov condition
\begin{equation}
\dfrac{\partial}{\partial x}\left[\left(\dfrac{\partial\mathcal{C}}{\partial
x}\right)^{-1}\dfrac{\partial}{\partial x}\left(\sigma\dfrac{\partial\mathcal{
C}}{\partial x}\right)\right]=0. 
\end{equation}
We now turn our attention to the form of the transforming function $\psi$. Using
equation~\eqref{eqn:dpsi_dx2} we obtain
\begin{equation}
\dfrac{\partial\ln\psi}{\partial x}=\dfrac{\zeta}{\sigma}\Rightarrow\psi(x,t)=
\psi(x_0,t)\exp\left(\zeta(t)\int_{x_0}^{x}\dfrac{1}{\sigma(x',t)}dx'\right),
\end{equation}
where $\psi(x_0,t)$ is an arbitrary function of time. With the explicit form of
$\zeta$ the previous expression can be conveniently rewritten as
\begin{equation}
\psi(x,t)=\psi(x_0,t)\dfrac{\left.\left(\sigma\dfrac{\partial\mathcal{C}}{
\partial x}\right)\right|_{(x_0,t)}}{\left.\left(\sigma\dfrac{\partial
\mathcal{C}}{\partial x}\right)\right|_{(x,t)}}.
\end{equation}
Since the form of $\psi(x_0,t)$ is arbitrary, we can set it equal to
\begin{equation}
\psi(x_0,t)=\left(\left.\left(\sigma\dfrac{\partial\mathcal{C}}{\partial x}
\right) \right|_{x_0,t} \right)^{-1},
\end{equation}
obtaining the result contained in the main text.\\

\section{Derivation of absorbing and reflecting propagators}
\label{appendix2}

We here provide a proof for equations \eqref{eqn:propagator_abs} and
\eqref{eqn:propagator_ref}. As mentioned in the main text, the functions
$\psi$ and $\tau$ map the original stochastic process onto the Wiener
process according to equation \eqref{eqn:SDE_solution}. Therefore, the problem
can be reduced to finding the propagator of the Wiener process with either
an absorbing or reflecting, time-dependent boundary. Interestingly, according
to equation \eqref{eqn:condition2}, after the space-time transformation the
boundary becomes
\begin{equation}
a(t)\to\psi(a(t),t)=v\tau+a_0,
\end{equation}
which is a linear function of the new time variable $\tau$. The propagator
$P_{W,a}$ of the Wiener process with absorbing linear-in-time boundary is
available in standard textbooks \cite{coxbook, goelbook} and reads
\begin{equation}
\label{eqn:wiener_abs}
P_{W,a}(y,\tau)=\dfrac{1}{\sqrt{2\pi\tau}}\left[\exp\left(-\dfrac{y^2}{2\tau}
\right)-\exp\left(-2a_0v-\dfrac{(y-2a_0)^2}{2\tau}\right)\right].
\end{equation}
Moreover, the relation between the propagator of $X_t$ and that of $W_{\tau}$
is given in equation \eqref{eqn:P_Pprime}; thus equation \eqref{eqn:propagator_abs} can
be obtained by multiplication by $\partial\psi/\partial x$ and substituting
$y\to\psi(x,t)$ in equation \eqref{eqn:wiener_abs}.

For the reflecting case we reason analogously---the propagator $P_{W,r}$ of
the Wiener process with reflecting boundary in $v\tau+a_0$ reads (see
\cite{coxbook,goelbook})
\begin{eqnarray}
\nonumber
P_{W,r}(y,\tau)&=&\left\{\frac{1}{\sqrt{2\pi\tau}}\left[\exp\left(-\frac{y^2}{
2\tau}\right)+\exp\left(-2a_0v-\frac{\left(y-2a_0\right)^2}{2\tau}\right)
\right]\right.\\
&&\left.-v\exp\left(-2v(y-a_0-v\tau)\right){\rm erfc}\left(\frac{y-2a_0-2v
\tau}{\sqrt{2 \tau}}\right)\right\},
\label{eqn:wiener_ref}
\end{eqnarray}
in terms of the complementary error function ${\rm erfc}$. The result,
equation \eqref{eqn:propagator_ref} in the main text, follows again by multiplication
by $\partial\psi/\partial x$ and substituting $y\to\psi(x,t)$.

We conclude this appendix with a short remark on the appropriate reflecting
boundary condition. This case is requires that the probability is conserved
over time. Therefore, if $x_0<a(t_0)$, integrating equation \eqref{eqn:FPE} in
the time-dependent domain $]-\infty,a(t)]$, we get
\begin{equation}
\int_{-\infty}^{a(t)}\dfrac{\partial P(x,t|x_0,t_0)}{\partial t}dx=-\int_{
-\infty}^{a(t)}\dfrac{\partial j(x,t|x_0,t_0)}{\partial x}dx.
\end{equation}
The left hand side can be manipulated as follows
\begin{eqnarray}
\nonumber
\int_{-\infty}^{a(t)}\dfrac{\partial P(x,t|x_0,t_0)}{\partial t}dx&=&\dfrac{
\partial}{\partial t}\int_{-\infty}^{a(t)}P(x,t|x_0,t_0)dx\\
&&-\dfrac{d a(t)}{dt}P(a(t),t|x_0,t_0)=- a'(t)P(x,t|x_0,t_0),
\end{eqnarray}
where we used the fact that $\int_{-\infty}^{a(t)}P(x,t|x_0,t_0)dx=1$.
Therefore, the boundary condition for the flux is not $j(a(t),t|x_0,t_0)=0$,
but rather
\begin{equation}
j(a(t),t|x_0,t_0)=\dfrac{d a(t)}{dt}P(a(t),t|x_0,t_0).
\end{equation}
Note that in the case of a constant boundary, i.e., $a'(t)=0$, the last
equation correctly states that the flux vanishes at the boundary. We could
not find this modified boundary condition in any standard reference of the
field.

\section{Didactic example: Ornstein-Uhlenbeck FPT and square-root boundaries
after space-time transformation}
\label{appendix_didactic}

In this appendix, we show how to apply step-by-step the technique developed
in the paper to the Ornstein-Uhlenbeck process (OUP), and how the FPTD of
the OUP for a constant boundary coincides with the FPTD of the Wiener process
for a square-root boundary. The SDE for the OUP reads
\begin{equation}
dX_t=-\dfrac{\kappa}{\gamma}X_tdt+\sqrt{\dfrac{2k_BT}{\gamma}}dW_t,
\end{equation}
where $\kappa$ is the stiffness of the harmonic potential. To be specific
we consider the constant parameters and $x_0=t_0=0$. From equation
\eqref{eqn:cherkasov} we obtain that the Cherkasov function in this case
reads
\begin{equation}
\mathcal{C}=\dfrac{\kappa}{\gamma},
\end{equation}
which does not depend either on $x$ or $t$, and thus fulfils condition
\eqref{eqn:condition}. We can thus apply our transformations; using
equation \eqref{eqn:psi} the space transformation reads
\begin{equation}
\psi(x,t)=\sqrt{\dfrac{\gamma}{2k_BT}}xe^{\kappa t/\gamma},
\end{equation}
while, using equation \eqref{eqn:tau}, the time transformation is
\begin{equation}
\tau(t)=\dfrac{\gamma}{2\kappa}\left(e^{2\kappa t/\gamma}-1\right).
\end{equation}
The propagator is immediately available substituting $\psi$ and $\tau$ into
equation \eqref{eqn:propagator}. 

Let us consider now the FPT for a constant threshold $a$. First, we need to
understand how the boundary changes, i.e.,
\begin{equation}
a\to\psi(a,t)=\sqrt{\dfrac{\gamma}{2k_BT}}ae^{\kappa t/\gamma}.
\end{equation}
Second, we need to reparametrise the transformed boundary with respect to the
new variable $\tau$,
\begin{equation}
\psi(a,\tau)=a\sqrt{\dfrac{\gamma}{2k_BT}}\sqrt{\dfrac{2\kappa}{\gamma}\tau+1},
\end{equation}
which is indeed a square-root function of $\tau$. Therefore, we proved the
equivalence between the FPTD we mentioned at the beginning of this appendix.

Using all the results available for the FPTD of the OU process
\cite{ricciardi1988,grebenkov}, we could, in principle, extend our results
to square-root boundaries, as well.


\section*{References}

\begin{thebibliography}{99}

\bibitem{Einstein} Einstein, A. {\"U}ber die von der molekularkinetischen
Theorie der W{\"a}rme geforderte Bewegung von in ruhenden Fl{\"u}ssigkeiten
suspendierten Teilchen. {\em Ann. Phys. (Leipzig)}. \textbf{322}, 549--560
(1905).

\bibitem{sutherland} Sutherland, W. LXXV. A dynamical theory of diffusion for
non-electrolytes and the molecular mass of albumin. {\em London, Edinburgh,
and Dublin Philosophical Magazine and Journal of Science}. \textbf{9},
781--785 (1905).

\bibitem{smoluchowski} von Smoluchowski, M. Zur kinetischen Theorie
der Brownschen Molekularbewegung und der Suspensionen. {\em Ann.
Phys. (Leipzig)}. \textbf{326}, 756--780 (1906).

\bibitem{langevin} Lemons, D. and Gythiel, A. Paul Langevin's 1908 paper "On
the theory of Brownian motion" ["Sur la th{\'e}orie du mouvement brownien",
C. R. Acad. Sci. (Paris) 146, 530-533 (1908)]. {\em Am. J. Phys.}. \textbf{65},
1079--1081 (1997).

\bibitem{wiener} Wiener, N. The average of an analytic functional and
the Brownian movement. {\em Proc. Natl. Acad. Sci. USA}. \textbf{7}, 294--298
(1921).

\bibitem{fokker} Fokker, A. Die mittlere Energie rotierender elektrischer
Dipole im Strahlungsfeld. {\em Ann. Phys. (Leipzig)}. \textbf{348}, 810--820
(1914).

\bibitem{planck} Planck, M. {\"U}ber einen Satz der statistischen Dynamik und
seine Erweiterung in der Quantentheorie. {\em Sitzungsberichte der Preussischen
Akademie der Wissenschaften zu Berlin}. \textbf{24}, 324--341 (1917).

\bibitem{kolmogorov1938} Kolmogorov, A. On the analytic methods of probability
theory, 1938, no.5. {\em Russian Math. Surv.}. \textbf{5}, 5--41 (1938).

\bibitem{levy} L{\'e}vy, P. Processus stochastiques et mouvement brownien
[Stochastic processes and Brownian motion] (Gauthier-Villars, Paris, 1948).

\bibitem{coffey} Coffey, W. T., Kalmykov, Yu. P., and Waldron, J. T.
The Lagevin equation (World Scientific, Singapore, 2005).

\bibitem{Ito1950} It\^{o}, K. Stochastic Differential Equations in a
Differentiable Manifold. {\em Nagoya Math. J.}. \textbf{1}, 35--47 (1950).

\bibitem{Hofling_2013} H{\"o}fling, F. and Franosch, T. Anomalous transport
in the crowded world of biological cells. {\em Rep. Prog. Phys.}. \textbf{76},
046602 (2013).

\bibitem{arbel-goren2023} Arbel-Goren, R., McKeithen-Mead, S., Voglmaier,
D., Afremov, I., Teza, G., Grossman, A. and Stavans, J. Target search by
an imported conjugative DNA element for a unique integration site along a
bacterial chromosome during horizontal gene transfer. {\em Nucleic Acids
Res.}. \textbf{51}, 3116--3129 (2023).

\bibitem{lene} Jeon, J.-H., Tejedor, V., Burov, S., Barkai, E., Selhuber-Unkel,
C., Berg-S{\o}rensen, K., Oddershede, L., and Metzler, R. In vivo anomalous
diffusion and weak ergodicity breaking of lipid granules. {\em Phys. Rev.
Lett.}. \textbf{106}, 048103 (2011).

\bibitem{garini} Bronstein, I., Israel, Y., Kepten, E., Mai, S., Shav-Tal,
Y., Barkai, E., and Garini, Y. Transient anomalous diffusion of telomeres
in the nucleus of mammalian cells. {\em Phys. Rev. Lett.}. \textbf{103},
018102 (2009).

\bibitem{jaenatc} Song, M. S., Moon, H. C., Jeon, J.-H., and Park, H. Y.
Neuronal messenger ribonucleoprotein transport follows an aging L{\'e}vy walk.
{\em Nature Comm.}. \textbf{9}, 344 (2018).

\bibitem{christine} Reverey, J. F., Jeon, J.-H., Bao, H., Leippe, M., Metzler,
R., and Selhuber-Unkel, C. Superdiffusion dominates intracellular particle
motion in the supercrowded space of pathogenic Acanthamoeba castellanii.
{\em Sci. Rep.}. \textbf{5}, 11690 (2015)

\bibitem{jaememb} Jeon, J.-H., Javanainen, M., Martinez-Seara,  H., Metzler, R.,
and Vattulainen, I. Protein crowding in lipid bilayers gives rise to non-Gaussian
anomalous lateral diffusion of phospholipids and proteins. {\em Phys. Rev. X}.
\textbf{6}, 021006 (2016).

\bibitem{mattijcp} Javanainen, M., Martinez-Seara, H., Metzler, R. and
Vattulainen, I. Diffusion of integral membrane proteins in protein-rich
membranes. {\em J. Phys. Chem. Lett.}. \textbf{8}, 4308 (2017).

\bibitem{animal_motion} Vilk, O., Aghion, E., Avgar, T., Beta, C., Nagel,
O., Sabri, A., Sarfati, R., Schwartz, D., Weiss, M., Krapf, D., Nathan, R.,
Metzler, R. and Assaf, M. Unravelling the origins of anomalous diffusion: From
molecules to migrating storks. {\em Phys. Rev. Res.}. \textbf{4}, 033055 (2022).

\bibitem{RevModPhys_active} Bechinger, C., Di Leonardo, R., L{\"o}wen, H.,
Reichhardt, C., Volpe, G. and Volpe, G. Active particles in complex and
crowded environments. {\em Rev. Mod. Phys.}. \textbf{88}, 045006 (2016).

\bibitem{BERKOWITZ2002861} Berkowitz, B. Characterizing flow and
transport in fractured geological media: A review. {\em Adv. Wat.
Res.}. \textbf{25}, 861--884 (2002).

\bibitem{Scher_1975} Scher, H. and Montroll, E. Anomalous transit-time
dispersion in amorphous solids. {\em Phys. Rev. B}. \textbf{12}, 2455--2477
(1975).

\bibitem{Bouchaud_Potters_2003} Bouchaud, J. and Potters, M. Theory of
Financial Risk and Derivative Pricing: From Statistical Physics to Risk
Management. (Cambridge University Press, Cambridge, UK, 2003).

\bibitem{brockmann_network} Brockmann, D. and Dirk Helbing The Hidden Geometry
of Complex, Network-Driven Contagion Phenomena. {\em Science}. \textbf{342},
1337--1342 (2013).

\bibitem{Redner_2001} Redner, S. A Guide to First-Passage Processes. (Cambridge
University Press, Cambridge, UK, 2001).

\bibitem{ralf_book_FPT} Metzler, R., Oshanin, G. and Redner, S. First-Passage
Phenomena and Their Applications. (World Scientific, Singapore, 2014).

\bibitem{aljaz} Godec, A. and Metzler, R. Universal proximity effect in target
search kinetics in the few encounter limit., {\em Phys. Rev. X}. \textbf{6},
041037 (2016).

\bibitem{denis} Grebenkov, D., Metzler, R. and Oshanin, G. Strong defocusing
of molecular reaction times: geometry and reaction control. {\em Commun. Chem.}.
\textbf{1}, 96 (2018).

\bibitem{aljaz1} Hartich, D., and Godec, A. Interlacing relaxation and
first-passage phenomena in reversible discrete and continuous space Markovian
dynamics. {\em J. Stat. Mech.}. \textbf{2019}, 024002 (2019).

\bibitem{kramers} Kramers, H. Brownian motion in a field of force and the
diffusion model of chemical reactions. {\em Physica}. \textbf{7}, 284--304
(1940).

\bibitem{hanggi_review} H{\"a}nggi, P., Talkner, P. and Borkovec,
M. Reaction-rate theory: fifty years after Kramers. {\em
Rev. Mod. Phys.}. \textbf{62}, 251--341 (1990).

\bibitem{Viswanathan} Viswanathan, G., Luz, M., Raposo, E. and Stanley,
H. The Physics of Foraging: An Introduction to Random Searches and Biological
Encounters. (Cambridge University Press, Cambridge, UK, 2011).

\bibitem{hufnagel_pnas} Hufnagel, L., Brockmann, D. and T. Geisel Forecast
and control of epidemics in a globalized world. {\em Proc. Natl. Acad. Sci.
USA}. \textbf{101}, 15124--15129 (2004).

\bibitem{Gross_2020} Gross, B., Zheng, Z., Liu, S., Chen, X., Sela, A., Li,
J., Li, D. and Havlin, S. Erratum: Spatio-temporal propagation of COVID-19
pandemics. {\em Europhys. Lett.}. \textbf{131}, 69901 (2020).

\bibitem{valenti_pre} Valenti, D., Fazio, G. and Spagnolo, B. Stabilizing
effect of volatility in financial markets. {\em Phys. Rev. E}. \textbf{97},
062307 (2018).

\bibitem{valenti_ijbc} Spagnolo, B. and Valenti, D. Volatility effects on
the escape time in financial market models. {\em Int. J. Bifurc. Chaos}.
\textbf{18}, 2775--2786 (2008).

\bibitem{Hartich_jpa} Hartich, D. and Godec, A. Extreme value statistics
of ergodic Markov processes from first passage times in the large deviation
limit. {\em J. Phys. A}. \textbf{52}, 244001 (2019).

\bibitem{Bray_majumdar} Bray, A. J., Majumdar, S. N., and Schehr,
G. Persistence and first-passage properties in nonequilibrium systems. {\em
Adv. Phys.}. {\bf 62} (3), 225-361 (2013).

\bibitem{target_search_problems} Grebenkov, D., Metzler, R. and Oshanin,
G. Target Search Problems. (Springer, Cham, CH, 2024).

\bibitem{cherkasov1957} Cherkasov, I. On the Transformation of the
Diffusion Process to a Wiener Process. {\em Theory Prob. Applic.}.
\textbf{2}, 373--377 (1957).

\bibitem{cherkasov_correction} Shirokov, F. On I. D. Cherkasov's Paper "On
the Transformation of a Diffusion Process to a Wiener Process".  {\em Theory
Prob. Applic.}. \textbf{9}, 156--157 (1964).

\bibitem{ricciardi1976} Ricciardi, L. On the transformation of diffusion
processes into the Wiener process. {\em J. Math. Anal. Applic.}. \textbf{54},
185--199 (1976).

\bibitem{ricciardi1984} Ricciardi, L., Sacerdote, L. and Sato, S. On an
integral equation for first-passage-time probability densities. {\em J.
Appl. Prob.}. \textbf{21}, 302--314 (1984).

\bibitem{wang} Wang, L. and P{\"o}tzelberger, K. Crossing Probabilities for
Diffusion Processes with Piecewise Continuous Boundaries. {\em Methodol.
Comput. Appl. Prob.}. \textbf{9}, 21--40 (2007).

\bibitem{Schmiedl} Schmiedl T. and Seifert U. Efficiency at
maximum power: An analytically solvable model for stochastic heat engines. {\em
Europhysics Letters} {\bf 81} 20003 (2007).

\bibitem{martinez} Mart\'inez, I. A., Rold\'an \'E., Dinis,
L., Petrov, D. and Rica, R. A. Adiabatic processes realized with a trapped
Brownian particle. {\em Phys. Rev. Lett.} {\bf 114}, 120601 (2015).

\bibitem{ricciardi1999} Ricciardi, L. M., Di Crescenzo,
A., Giorno, V. and Nobile A. G. An outline of theoretical and algorithmic
approaches to first passage time problems with applications to biological
modeling. {\em Mathematica Japonica} {\bf 50}, 247--322 (1999).

\bibitem{oksendal} {\O}ksendal, B. Stochastic differential equations.
(Springer, Berlin, 2003).

\bibitem{martingale_review} Rold{\'a}n {\'E}., Neri, I., Chetrite, R., Gupta, S.,
Pigolotti, S., J{\"u}licher, F. and Sekimoto, K. Martingales for physicists:
a treatise on stochastic thermodynamics and beyond. {\em Adv. Phys.}.
\textbf{72}, 1--258 (2023).

\bibitem{bray_log} Bray, A. J. Random walks in logarithmic and power-law potentials, nonuniversal persistence, and vortex dynamics in the two-dimensional XY model. {\em Phys. Rev. E} {\bf 62}, 103--112 (2000).

\bibitem{giampaoli_anharmonic} Giampaoli, J. A., Strier, D. E., Batista, C., Drazer, G. and Wio, H. S. Exact expression for the diffusion propagator in a family of time-dependent anharmonic potentials, {\em Phys. Rev. E} {\bf 60}, 2540 (1999).

\bibitem{ryabov_log-harmonic} Ryabov, A., Dierl, M., Chvosta,
P., Einax, M. and Maass, P. Work distribution in a time-dependent
logarithmic-harmonic potential: exact results and asymptotic analysis, {\em J. Phys. A: Math. Theor.} {\bf 46} 075002.

\bibitem{skiadas} Skiadas, C. H. Exact solutions of stochastic
differential equations: Gompertz, generalized logistic and revised exponential,
{\em Methodol. Comput. Appl. Probab.} {\bf 12}:261--270 (2010).

\bibitem{tubikanec} Tubikanec, I., Tamborrino, M., Lansky, P. and Buckwar, E. Qualitative properties of different numerical methods for the
inhomogeneous geometric Brownian motion, {\em Journal of Computational and Applied Mathematics} {\bf 406} (2022) 113951.

\bibitem{molini} Molini, A., Talkner, P., Katul, G. and Porporato,
A. First passage time statistics of Brownian motion with purely time
dependent drift and diffusion. {\em Physica A}. \textbf{390}, 1841--1852 (2011).

\bibitem{grebenkov} Grebenkov, D. First exit times of harmonically trapped
particles: a didactic review. {\em J. Phys. A}. \textbf{48}, 013001 (2014).

\bibitem{liptser2013statistics} Liptser, R. and Shiryaev, A. Statistics of
random processes: I. General theory. (Springer, Berlin, 2013).

\bibitem{merton1971theory} Merton, R. Theory of rational option pricing. {\em
Bell J. Econom. Manag. Sci.}. \textbf{4}, 141--183 (1971).

\bibitem{robbins_siegmund1970} Robbins, H. and Siegmund, D. Boundary Crossing
Probabilities for the Wiener Process and Sample Sums. {\em Ann. Math. Stat.}.
\textbf{41}, 1410--1429 (1970).

\bibitem{siegmund1986} Siegmund, D. Boundary Crossing Probabilities and
Statistical Applications. {\em Ann. Statist.}. \textbf{14}, 361--404 (1986).

\bibitem{durbin} Durbin, J. and Williams, D. The first-passage density of
the Brownian motion process to a curved boundary. {\em J. Appl. Probab.}.
\textbf{29}, 291--304 (1992).

\bibitem{alili2010} Alili, L. and Patie, P. On the first crossing times of
a Brownian motion and a family of continuous curves. {\em C. R. Math.}.
\textbf{340}, 225--228 (2005).

\bibitem{lerche} Lerche, H. Boundary crossing of Brownian motion. (Springer,
New York, NY, 1986)

\bibitem{wang97} Wang, L. and P{\"o}tzelberger, K. Boundary
crossing probability for Brownian motion and general boundaries, {\em
J. Appl. Prob.} {\bf 34} 54-65 (1997).

\bibitem{coxbook} Cox, D. The theory of stochastic processes. (Chapman \&
Hall/CRC, London, UK, 1977).

\bibitem{goelbook} Goel, N. and Richter-Dyn, N. Stochastic models in
biology. (Elsevier, Amsterdam, 1975).

\bibitem{breiman} Breiman, L. First exit times from a square root
boundary. {\em Proc. Fifth Berkeley Symposium on Mathematical Statistics and
Probability, Volume II, Part II}. 9--16 (1967).

\bibitem{ciliberto2017experiments} Ciliberto, S. Experiments in stochastic
thermodynamics: Short history and perspectives. {\em Phys. Rev. X}.
\textbf{7}, 021051 (2017).

\bibitem{martinez2017colloidal} Mart\'inez, I., Rold{\'a}n ,{E}., Dinis, L. and
Rica, R. Colloidal heat engines: a review. {\em Soft Matter}. \textbf{13},
22--36 (2017).

\bibitem{frydel} Frydel, D. Statistical mechanics of passive Brownian particles
in a fluctuating harmonic trap. {\em Phys. Rev. E}. \textbf{110}, 024613 (2024).

\bibitem{siegert} Siegert, A. J. F. On the First Passage Time Probability
Problem. {\em Phys. Rev.}. \textbf{81}, 617--623 (1951).

\bibitem{ricciardi1977} Ricciardi, L. Diffusion processes and related topics
in biology. (Springer, Berlin, 1977).

\bibitem{garrahan_pre} Garrahan, J. P. and Ritort,
F. Generalized continuous Maxwell demons. {\em Phys. Rev. E} {\bf 107} 034101
(2023).

\bibitem{saha_prl} Saha, T. K., Lucero J. N. E., Ehrich, J.,
Sivak, D. A. and Bechhoefer, J. Bayesian information engine that optimally
exploits noisy measurements. {\em Phys. Rev. Lett.} {\bf 129} 130601 (2022).

\bibitem{tohme} Tohme, T., Bedoya, V., Di Bello, C.,
Bresque L., Manzano, G., Rold\'an, E. Gambling Carnot engine. E-print
arXiv:2409.17212.

\bibitem{archambault_epl} Archambault, A., Crauste-Thibierge,
C., Ciliberto, S. and Bellon, L. Inertial effects in discrete sampling
information engines. {\em Europhysics Letters} {\bf 148} 41002 (2024).  

\bibitem{archambault_arxiv} Archambault, A., Crauste-Thibierge,
C., Imparato, A., Jarzynski, C., Ciliberto, S. and Bellon, L. Information
engine fueled by first-passage times. E-print arXiv:2407.17414.

\bibitem{martinez2015adiabatic} Mart\'inez, I., Rold{\'a}n, {\'E}., Dinis, L.,
Petrov, D. and Rica, R. Adiabatic processes realized with a trapped Brownian
particle. {\em Phys. Rev. Lett.}. \textbf{114}, 120601 (2015).

\bibitem{polyanin} Polyanin, A. and Nazaikinskii, V. Handbook of Linear
Partial Differential Equations for Engineers and Scientists. (Chapman,
London, UK, 2015).

\bibitem{capocelli1974} Capocelli, R. M. and Ricciardi,
L. M. Growth with regulation in random environment. {\em Kybernetik} {\bf 15},
147--157 (1974).

\bibitem{gompertz} Gompertz, B. On the nature of the function 
expressive of the law of human mortality, and on the mode of determining the
value of life contingencies. {\em Philos. Trans. R. Soc.}  \textbf{115},
513--585 (1825).

\bibitem{laird} Laird, A. K. Dynamics of tumor growth.
{\em British J. Cancer} {\bf 18}, 490--502 (1964). 

\bibitem{loos} Loos, S. A. M. and Klapp, S. H. L.
Fokker-Planck equations for time-delayed systems via Markovian embedding.
{\em J. Stat. Phys.} {\bf 177}, 95--118 (2019).

\bibitem{de2021generating} De Bruyne, B., Majumdar, S.,
Orland, H. and Schehr, G. Generating stochastic trajectories with global
dynamical constraints. {\em J. Stat. Mech.}. \textbf{2021}, 123204 (2021).

\bibitem{chubynsky_DD} Chubynsky, M. and Slater, G. Diffusing
Diffusivity: A Model for Anomalous, yet Brownian, Diffusion. {\em
Phys. Rev. Lett.}. \textbf{113}, 098302 (2014).

\bibitem{jain} Jain, R. and Sebastian, K. L. Diffusing diffusivity: survival
in a crowded rearranging and bounded domain. {\em J. Phys. Chem. B}.
\textbf{120}, 9215--9222 (2016).

\bibitem{Sposini_2020} Sposini, V., Grebenkov, D., Metzler, R., Oshanin, G. and
Seno, F. Universal spectral features of different classes of random-diffusivity
processes. {\em New J. Phys.}. \textbf{22}, 063056 (2020).

\bibitem{cocconi_ou2} Cocconi, L., Alston, H., Romano, J. and Bertrand, T. The
OU2 process: characterising dissipative confinement in noisy traps. {\em
New J. Phys.}. \textbf{26}, 103016 (2024).

\bibitem{bechhoefer2021control} Bechhoefer, J. Control theory for
physicists. (Cambridge University Press, Cambridge, UK, 2021).

\bibitem{khodabandehlou2024vanishing} Khodabandehlou, F., Maes, C.,
Maes, I. and Neto\v{c}n\`{y}, K. The vanishing of excess heat for
nonequilibrium processes reaching zero ambient temperature. {\em Ann.
Henri Poincar{\'e}}. \textbf{25} 3371-3403 (2024).

\bibitem{kolchinsky2021work} Kolchinsky, A. and Wolpert, D. Work, Entropy
production, and thermodynamics of information under protocol constraints. {\em
Phys. Rev. X}. \textbf{11}, 041024 (2021).

\bibitem{lamperti} Lamperti, J. Semi-stable stochastic processes. {\em
Trans. Amer. Math. Soc.}. \textbf{104} 62--78 (1962).

\bibitem{Mao_book} Mao, X. Stochastic differential equations
and applications, second ed., pp. 91--106 (Woodhead Publishing, Sawston, UK,
2011).

\bibitem{ricciardi1988} Ricciardi, L. and Sato, S. First-passage-time
density and moments of the Ornstein-Uhlenbeck process. {\em J. Appl.
Prob.}. \textbf{25}, 43--57 (1988).

\end{thebibliography}
\end{document}